\documentclass[12pt]{iopart}

\usepackage{iopams}
\usepackage{graphicx,amssymb}
\usepackage{color}
\newtheorem{lemma}{Lemma}
\newtheorem{proposition}{Proposition}
\newtheorem{theorem}{Theorem}

\newtheorem{definition}{Definition}

\newcommand{\be}{\begin{equation}}
\newcommand{\ee}{\end{equation}}
\def\dif{{\rm d}}

\begin{document}
\title[Perfect fluid flows]
{On the flow of perfect energy tensors}

\author{Juan Antonio S\'aez$^1$, Salvador Mengual$^{2}$\footnote{Author to whom any correspondence should be adressed.} and Joan Josep Ferrando$^{2,3}$}

\address{$^1$\ Departament de Matem\`atiques per a l'Economia i l'Empresa,
Universitat de Val\`encia, E-46022 Val\`encia, Spain}

\address{$^2$\ Departament d'Astronomia i Astrof\'{\i}sica, Universitat
de Val\`encia, E-46100 Burjassot, Val\`encia, Spain}

\address{$^3$\ Observatori Astron\`omic, Universitat
de Val\`encia, E-46980 Paterna, Val\`encia, Spain}

\ead{juan.a.saez@uv.es; salvador.Mengual@uv.es; joan.ferrando@uv.es}

\begin{abstract}
The necessary and sufficient conditions are obtained for a unit time-like vector field $u$ to be the unit velocity of a divergence-free perfect fluid energy tensor. This plainly kinematic description of a conservative perfect fluid requires considering eighteen classes defined by differential concomitants of $u$. For each of these classes, we get the additional constraints that label the flow of a conservative energy tensor, and we obtain the pairs of functions $\{\rho,p\}$, energy density and pressure, which complete a solution to the conservation equations.
\end{abstract}
%

%



\section{Introduction}
\label{sec-intro}


In Relativity, the energetic description of the evolution of a perfect fluid is
given by its energy tensor $ T = (\rho+ p) u \otimes u + p \, g$,
where $\rho$, $p$ and $u$ are, respectively, the {\em energy
density}, the {\em pressure} and {\em unit velocity} of the fluid, and $g$ is the spacetime metric. Every perfect fluid energy tensor $T \equiv \{u, \rho, p\}$ defines a point of the space ${\bf U} \times {\bf F} \times {\bf F}$, where ${\bf U}$ denotes the set of the time-like unit vector fields, and ${\bf F}$ the set of  functions over a given domain of the spacetime. 
For isolated media, we must impose the conservation of the energy tensor, 
\begin{equation}
\label{conservation-1}  \nabla \cdot T = 0 \, ,
\end{equation}
a first order differential system of four equations for the {\em hydrodynamic quantities} $\{u, \rho, p\}$. The set of solutions of (\ref{conservation-1}), ${\cal S} = \{T \equiv (u, \rho, p) \,|\,  \nabla \cdot T = 0 \}$, defines a paralle\-lepiped ${\cal S} = {\bf U}_c \times {\bf F}_{\rho} \times {\bf F}_p \subset {\bf U} \times {\bf F} \times {\bf F}$. 

Note that ${\bf U}_c$ is a non-empty set. A simple example is given by a unit vector $u$ defining a geodesic, irrotational and non-expanding flow. In this case $u =- \dif \tau$, and if $p=p(\tau)$ and $\rho$ is a $u$-invariant function, then the triad $\{u, \rho,p\}$ is a solution of (\ref{conservation-1}), and thus $u \in {\bf U}_c$.

Furthermore, ${\bf U}_c$ is a proper subset of ${\bf U}$, ${\bf U}_c \not={\bf U}$. This fact will be thoroughly analyzed in this paper. Now, we offer an example. Let us consider inertial spherical coordinates $(t,r,\vartheta, \varphi)$ in the Minkowski spacetime, and define the unit vector $u= (\cosh \psi(\varphi), 0, r^{-1} \sinh \psi(\varphi), 0) \in {\bf U}$, where $\psi= \psi(\varphi)$ is a non-constant arbitrary real function. A straightforward calculation shows that the conservation equation (\ref{conservation-1}) then becomes a differential system for the functions $\{\rho,p\}$ that does not admit solution. Thus, $u \notin {\bf U}_c$.

Consequently, in general, a conservative perfect energy tensor having an arbitrary flow does not exist. Thus, it is natural to ask the following question: is it possible to intrinsically define ${\bf U}_c$? Or, more precisely, is it possible to express, solely in terms of $u$ and its derivatives, the necessary and sufficient conditions for $u$ to be the unit velocity of a conservative perfect energy tensor? The main goal of this paper is to show that the answer is affirmative, and to obtain these conditions.

Our results mean that the system (\ref{conservation-1}) for the hydrodynamic quantities $\{u, \rho, p\}$ admits a {\em conditional system} for the sole {\em kinematic quantity} $u$:
\be  \label{D(u)}
{\cal D}(u) =0 \, .
\ee
This means that (\ref{D(u)}) is a consequence of (\ref{conservation-1}), and conversely, for any solution $u$ of (\ref{D(u)}), a solution $\{u, \rho, p\}$ of (\ref{conservation-1}) exists. In other words, (\ref{D(u)}) is the integrability condition for the system (\ref{conservation-1}) to admit a solution $\{\rho,p\}$.

Consequently, our study shows two aspects that we will analyze in depth: (i) the {\em direct problem}, namely, the determination of the conditional system (\ref{D(u)}) from the initial one (\ref{conservation-1}), and (ii) the {\em inverse problem}, namely, the obtention of the solutions of (\ref{conservation-1}) associated with a given solution of (\ref{D(u)}).

The search for the conditional system (\ref{D(u)}) leads to a classification of the time-like unit vectors in eighteen classes. For each class, we obtain the necessary and sufficient conditions in $u$ to ensure that it belongs to ${\bf U}_c$ (direct problem). Furthermore, for each class we solve the inverse problem by obtaining the pairs $\{\rho,p\}$ that complete a solution to the system (\ref{conservation-1}).

It should be mentioned that a similar approach was taken years ago for {\em the system of the barotropic hydrodynamics} \cite{CF-barotrop}, and more recently for {\em the system of the classical ideal gas hydrodynamics} \cite{FS-KCIG}. In this relativistic hydrodynamics framework, the direct and inverse problems have been analyzed for the thermodynamic perfect fluids \cite{Coll-Ferrando-termo} \cite{CFS-LTE}, for a generic ideal gas \cite{CFS-LTE}, for a classical ideal gas \cite{CFS-CIG} and, not long ago, for a relativistic Synge gas \cite{MFS-Synge}. These studies have proven useful in interpreting some families of perfect fluid solutions to the Einstein equations \cite{CFS-CIG, MFS-Synge, C-F, FS-SS, CFS-parabolic,CFS-regular,CFS-CC, FM-T1, FM-T2, MF-LT}.

Furthermore, the study of conditional systems associated with a differential system and the analysis of the corresponding direct and inverse problems have proven useful in other contexts. For example, the Rainich \cite{rainich} theory for the non-null electromagnetic field precisely consists in
obtaining the conditional system (involving the sole metric tensor) associated to the Einstein-Maxwell  equations. Still in the electromagnetic framework, we can quote the interpretation of the Teukolsky-Press relations \cite{teukolsky-press} \cite{CF-TP}, and in a more formal context the study of the Rainich approach to the Killing-Yano tensors \cite{fsKY} and to the Killing and conformal tensors \cite{cfs-KT}. 

The IDEAL (intrinsic, deductive, explicit and algorithmic) characterization of a family of metrics can also be formally identified as the answer to a direct problem (see \cite{fswarped, fs-SSST, FS-SS, Khavkine, Khavkine-b} and references therein).

When the fluid is the source of the gravitational field, the Ricci and Bianchi identities relate the Riemann tensor and its derivatives to those of the fluid kinematic quantities. Thus, the formalism for studying the metric equivalence developed by Cartan \cite{cartan,Karlhede} based on the derivatives of the Riemann tensor induces a classification of the perfect fluid solutions that constrains the kinematic quantities and their derivatives. This is the approach that can be found in \cite{BRADLEY} and it is illustrated by studying the conformally flat perfect fluid solutions. As a sample of the constraints that the field equations impose on the fluid flow we can quote the following result by Ellis \cite{Ellis1967}: a shear-free dust solution must be either expansion-free or irrotational. Later, it was conjectured that this property also holds if we change the dust condition to a barotropic equation of state \cite{Treciokas}, and it has been shown in many particular cases (see \cite{Sopuerta, VandenBergh1999, Slobodeanu, VandenBergh2016} and references therein).

Many perfect fluid solutions to Einstein's equations have been obtained by imposing purely kinematic conditions a priori. Thus, the geodesic, irrotational, shear-free or non-expanding character of the flow plays an important role when integrating the field equations for the spherically symmetric case (see \cite{kramer} and references therein). Another approach in finding new solutions is to restrict the Petrov-Bel type of the spacetime, and then the flow of the fluid is usually constrained by conditions involving the Weyl tensor. 

Perfect fluid solutions of Petrov-Bel type D with irrotational and non-shearing flow were studied by Barnes \cite{Barnes}. The unit velocity is then aligned with the Weyl principal plane, a property that the family of solutions classified  and studied in \cite{WAINWRIGHT, CARMINATI-WAINWRIGHT} also keep. More recently, the complete classification of aligned type D purely magnetic perfect fluids \cite{VANDENBERGH-WYLLEMAN} and of purely magnetic, irrotational, geodesic perfect fluids \cite{WYLLEMAN-VANDENBERGH} has been achieved. Other solutions obtained by also imposing constraints on both the Weyl tensor and the fluid flow can be found in \cite{Bas-Wylleman, VdenBergh-Wylleman-2009}.

All these classifications and studies exploit the relations of the fluid with the curvature tensor via Einstein's equations. In contrast, here we perform a different approach by classifying the fluids with independence of being or not the source of the gravitational field and just taking into account that the energy tensor is divergence-free, obtaining then an exhaustive and purely kinematic classification.

This paper is organized as follows. In section \ref{sec-opening}, we present some preliminary results on the direct and inverse problems. We analyze the case of a geodesic flow and we give a general result that applies for the non-geodesic case. 

The study of the non-geodesic rotating flows leads to numerous classes that will be considered in three sections: in section \ref{sec-wnot=0_a}, we analyze general features of this case and show that a basis of vectors, concomitants of $u$, can be defined to examine the stated problem; section \ref{sec-wnot=0_b} (respectively, section \ref{sec-wnot=0_c}) is devoted to investigating the non-geodesic rotating flows with a rotating (respectively, irrotational) acceleration vector. In section \ref{sec-w=0}, we analyze the classes with a non-geodesic irrotational flow.

In section \ref{sec-quadres}, we summarize the main results of the paper in an enlightening form. We present three tables. The first one offers the scalar and vectorial $u$-concomitants needed in order to define the classes and to characterize the flows in {\bf U}$_c$. The second one presents the conditions that define the eighteen classes of unit vectors. The third one gives, for every class, the necessary and sufficient conditions characterizing the velocities in {\bf U}$_c$, and the corresponding solution to the inverse problem. We also present a flow diagram with an algorithm enabling us to distinguish every class.

In section \ref{sec-examples}, we present some examples that illustrate the interest of our results and comment on further extensions and applications in progress: stationary flows, radial infall onto a central object, stationary circular flows in a stationary axisymmetric spacetime, and the flow of the Stephani Universes. 

Finally, in section \ref{sec-discussion}, we compare our approach with the usual kinematic classifications, and comment on the further work that our results suggest. A subsection is devoted to outlining a subject that is work in progress: the analyses of the thermodynamic behavior of the fluids of each one of the defined classes. 

In this paper, we work on an oriented spacetime with a metric tensor $g$ of signature $\{-,+,+,+\}$. For the metric product of two vectors, we write $(x,y) = g(x,y)$ and $x^2 = g(x,x)$. The symbols $\nabla$, $\nabla \cdot$, $\dif$, $\wedge$ and $*$ denote, respectively, the covariant derivative, the divergence operator, the exterior derivative, the exterior product and the Hodge dual operator, while $i(x)t$ stands for the interior product of a vector field $x$ and a p-tensor $t$. We shall denote with the same symbol a tensor and its associated tensors by raising and lowering indexes with the metric $g$.


\section{Conservation equations and opening results}
\label{sec-opening}

An energy tensor of the form $T = (\rho+ p) u \otimes u + p \, g$, with $\rho+p \not=0$, that fulfills the divergence-free condition (\ref{conservation-1}) is called a {\em perfect energy tensor}. In terms of the hydrodynamic quantities $\{u, \rho,p\}$, the conservation equation (\ref{conservation-1}) takes the expression:
\begin{eqnarray}
\dif p  + \dot{p} u + (\rho + p) a = 0 \, ,  \label{con-eq1} \\[2mm]
\dot{\rho} + (\rho+ p) \theta = 0 \, ,   \label{con-eq2}
\end{eqnarray}
where $a$ and $\theta$ are, respectively, the acceleration and the
expansion of $u$, 
\be \label{acceleracio}
a=a[u] \equiv i(u) \nabla u \, , \qquad  \theta=\theta[u] \equiv \nabla \cdot u  \, ,
\ee
and where a dot denotes the directional
derivative, with respect to $u$, of a quantity $q$, $\dot{q} = u(q)
= u^{\alpha} \partial_{\alpha} q$. From now on, we write $h=h[u]$ to indicate that $h$ is a (tensorial) differential concomitant of the vector $u$. We will use Greek letters for scalar concomitants and lowercase Latin letters for vectorial concomitants. For instance, we denote $w$ and $v$ the rotation vectors associated with the unit velocity $u$ and the acceleration vector, respectively,
\be \label{rotations}
w = w[u] \equiv *(u \wedge \dif u) \, , \qquad  v = v[u] \equiv *(a \wedge \dif a)  \, .
\ee

Note that the change of the scalar quantities $\rho$ and $p$ by a constant factor, and by an additive constant without changing $\rho+p$, leaves equations (\ref{con-eq1}-\ref{con-eq2}) invariant. Consequently, we have the following general result concerning the inverse problem:
\begin{proposition} \label{propo-uf-rop}
If $\{u, \rho,p\}$ is a solution of the conservation equations {\em (\ref{con-eq1}-\ref{con-eq2})}, then $\{u, \bar{\rho},\bar{p}\}$ is also a solution, where
\be \label{kappa_i}
\bar{p} = c_1 p + c_2 \, , \qquad  \bar{\rho} = c_1 \rho - c_2  \, ,
\ee
$c_1, c_2$ being two arbitrary constants.
\end{proposition}

As previously commented in the introduction, obtainig the conditional system (\ref{D(u)}) associated to the differential system (\ref{con-eq1}-\ref{con-eq2}) requires considering different cases that lead to a classification of the perfect fluid flows.


\subsection{Geodesic flows: $a=0$ {\rm (C$_1$)}}
\label{subsec-a=0}

If $u$ defines a geodesic flow, then equation (\ref{con-eq1}) becomes $\dif p = - \dot{p} u$. Then, either $u$ is an integrable 1-form or $p$ is a constant. In the first case we have an irrotational flow, $u$ must be closed, and a function $\tau$ exists such that $u = - \dif \tau$. Consequently, we can take $p \equiv p(\tau)$ as an arbitrary function, and equation (\ref{con-eq2}) becomes a linear differential equation for $\rho$, which can be integrated. The general solution depends on an arbitrary u-invariant function, $\psi = \psi(x^1, x^2, x^3)$ (with $(x^i)$ defining a coordinate system in the hypersurface $\tau = constant$), and it is given by 
\begin{equation} \label{rho-a=0-1} 
\rho = \left[ \psi(x^i)
-  \int \! \theta \, p(\tau) \,  e^{\int \! \theta \dif \tau}  \ \dif \tau\right]
 e^{- \int \!{\theta \dif \tau}} \, .
\end{equation}
If we have a rotating flow, then necessarily $p=p_0$. Therefore, the solution of equation (\ref{con-eq2}) is of the form
\begin{equation} \label{rho-a=0-2} 
\rho = - p_0 +  \psi(x^i) e^{- \int \!{\theta \dif \tau}} \, ,
\end{equation}
where $\psi = \psi(x^1, x^2, x^3)$ is an arbitrary u-invariant function, $(\tau, x^i)$ being a coordinate system adapted to the vector field $u$, $u = \partial_{\tau}$. Consequently, we arrive to the following result which solves the direct and inverse problems for a geodesic flow:
\begin{proposition}  \label{prop-a=0}
A  geodesic time-like unit vector $u$ is always the velocity of a perfect energy tensor.

If $u$ is irrotational ($w=0$), then $u = -\dif \tau$, and the pressure is given by an arbitrary real function $p=p(\tau)$ and the energy density $\rho$ is given by {\em (\ref{rho-a=0-1})}. If $u$ defines a rotating flow ($w\not=0$) the pressure is an arbitrary constant $p=p_0$ and the energy density $\rho$ is given by {\em (\ref{rho-a=0-2})}. In both cases $\psi(x^i)$ is an arbitrary u-invariant function.
\end{proposition}

From here on, we suppose that $u$ defines a non-geodesic flow.


\subsection{A general result on non-geodesic flows: $a \not=0$}
\label{subsec-X}

Equation (\ref{con-eq1}) can be written as 
\begin{equation} \label{s-1}
s = - \frac{\dif p}{\rho + p} \, , \qquad \quad s \equiv  a  + \Psi u \, , \qquad \Psi \equiv \frac{\dot{p}}{\rho +p} \, .
\end{equation}

Taking the differential of $s$ we obtain
\begin{equation} \label{ds}
\dif s = q \wedge s \, ,  \qquad \dif{q} =0 \, , \qquad  q \equiv  - \dif \ln(\rho +p) \, .
\end{equation}
Then, making the inner product with $u$ of the first equation and using
(\ref{con-eq2}) to eliminate $\dot{\rho}$, we get:
\be  \label{uds}
i(u) \dif{s}= - \Psi s + \theta s + \Psi q \, .
\ee
We can summarize these results as follows. 
\begin{lemma} \label{lema1}
If $u$ is a non-geodesic velocity of a perfect energy tensor, then a function $\Psi$ exists such that the pair $\{u, \Psi\}$ fulfills equations 
\be \label{eq-uchi}
\dif s = q \wedge s\, ,  \quad \dif q =0 \, , \quad s = a + \Psi u \, , \quad
\Psi q = (\Psi - \theta ) s + i(u) \dif s \, .
\ee
\end{lemma}
Note that $s$ depends on $u$ and $\Psi$. When $\Psi \not=0$, the last equation in (\ref{eq-uchi}) means that $q$ is also determined by $\Psi$ and $u$. When $\Psi=0$ (equivalent to $\dot{p}=0$) we cannot compute $q$ from this equation.

Now, we can prove the converse of the lemma above as follows. Let us suppose that for a non-geodesic vector $u$ a function $\Psi$ exists such that the pair $\{u, \Psi\}$ fulfills equations (\ref{eq-uchi}). Since $a\neq 0$, we have $s \neq 0$. The two first conditions in equation (\ref{eq-uchi}) mean that $s$ is an integrable 1-form and that $q$ is closed (or zero). Then, two functions $\lambda, \mu$ exist such that
\begin{equation} \label{s-q}
 s = e^{\mu} \dif \lambda \, , \qquad \qquad q = \dif \mu \, .
\end{equation}
These relations and the two last equations in (\ref{eq-uchi}) imply that
\begin{equation} \label{new1}
\theta - \Psi = (u, q) = \dot{\mu} \, , \qquad  \Psi = -  (u,s) \ = - e^{\mu} \dot{\lambda} \, .
\end{equation}
Then, from these relations and (\ref{eq-uchi}) we can easily prove that functions
\begin{equation} \label{p-rho}
p = - \lambda \, , \qquad \quad \rho = \lambda + e^{-\mu} \, , 
\end{equation}
complete a solution $\{u, \rho, p\}$ to the conservation system (\ref{con-eq1}-\ref{con-eq2}). Consequently, we can state:
\begin{proposition} \label{prop-anot=0}
A non-geodesic time-like unit vector $u$ is the velocity of a perfect energy tensor if, and only if, a function $\Psi$ exists such that the pair $\{u, \Psi\}$ fulfills equations {\em (\ref{eq-uchi})}.

Moreover, for each solution $\{u, \Psi\}$ to {\em (\ref{eq-uchi})}, the energy density $\rho$ and the pressure $p$ are constrained by
\be \label{uchi-rhop}
\dif p =  - (\rho +p)s , \qquad \quad \dif \ln (\rho +p) =-q \, .
\ee
\end{proposition}

In the proof of the sufficient condition of the above proposition, we have shown that equations (\ref{uchi-rhop}) admit, at least, a solution $\{\rho, p\}$ when conditions (\ref{eq-uchi}) hold. Solving the inverse problem requires determining all the solutions $\{\rho, p\}$. When $\Psi \not=0$, $q$ is fixed by $\Psi$ and $u$, and then functions $\lambda, \mu$ are determined by (\ref{s-q}) up to additive constants. Nevertheless, when $\Psi=0$ the vector $q$ is not fixed, and only the first equation in (\ref{s-q}) constrains functions $\lambda, \mu$, that is, they depend on an arbitrary function of $\lambda$. Thus, we have the following results on the inverse problem:
\begin{proposition} \label{prop-anot=0-inverse}
Let $\{u, \Psi\}$ be a solution to equations {\em (\ref{eq-uchi})}, where $u$ is a non-geodesic time-like unit vector, and let $\{\lambda, \mu\}$ be two functions fulfilling {\em (\ref{s-q})}.  Then, the solutions $\{u, \bar{\rho},\bar{p}\}$ of the conservation equations {\em (\ref{con-eq1}-\ref{con-eq2})} are defined by {\em (\ref{kappa_i})}, where $\{\rho,p\}$ are given by
\begin{eqnarray}
p = - \lambda, \ \quad \rho = -p + e^{-\mu}  \qquad  & \quad {\rm  if} \quad  \Psi \not=0 \, , \label{rhop-not0} \\
p = p(\lambda),  \quad \rho = -p(\lambda) - p'(\lambda) e^{-\mu}  \quad  & \quad {\rm  if} \quad  \Psi =0 \, .\label{rhop-0}
\end{eqnarray}
\end{proposition}

The case $\Psi=0$ corresponds to an isobaric evolution ($\dot{p} =0$). Then, $s=a$, and equations (\ref{eq-uchi}) imply
\be \label{eq-chi=0}
\dif a \wedge a =0 \,, \qquad \quad  i(u) \dif a = \theta \, a \, .
\ee
Note that the first equation above means that $a = e^{\mu} \dif \lambda$, and then $da = q \wedge a$ with $q = \dif \mu$, and equations (\ref{eq-uchi}) hold. Thus, we have obtained the following result:
\begin{proposition} \label{prop-chi=0}
A non-geodesic time-like unit vector $u$ is the velocity of a perfect energy tensor with $\dot{p}=0$ if, and only if, it fulfills equations {\em (\ref{eq-chi=0})}. 

Then, $a = e^{\mu} \dif \lambda$, and the isobaric solutions $\{u, \bar{\rho},\bar{p}\}$ of the conservation equations {\em (\ref{con-eq1}-\ref{con-eq2})} are defined by {\em (\ref{kappa_i})}, where $\{\rho,p\}$ are given in {\em(\ref{rhop-0})}.
\end{proposition}
It is worth remarking that the above proposition states that a unit time-like vector $u$ fulfilling (\ref{eq-chi=0}) is the velocity of an isobaric solution. Nevertheless, this $u$ can also be the velocity of a non-isobaric evolution when another pair $\{u, \Psi\}$, with $\Psi \not=0$, is also a solution of equations (\ref{eq-uchi}). Some of the classes with a non-geodesic flow considered hereon are compatible with equation (\ref{eq-chi=0}), and this constraint defines a subset of flows in each of these classes.


\subsection{Case $(w,a) \not=0$ {\em (C$_2$)}}
\label{subsec-(wa)not=0}


Now, we study the more regular case. From (\ref{eq-uchi}) we obtain  $\dif s  = \dif a + \dif \Psi \wedge u + \Psi \dif u = q \wedge s$. Then, if we make the exterior product of this equation with $u\wedge a$, we get
\be
 \Psi \, \dif u \wedge u \wedge a + \dif a \wedge u \wedge a=0 \, ,
\ee
and taking the dual operator we obtain $(w, a) \Psi = (v, u)$.  So, if $u$ is a vector satisfying $(w,a) \not=0$, we can compute the function $\Psi$ in terms of $u$:
 \begin{equation} \label{chi-C2}
 \Psi = \Psi_2[u] \equiv  \frac{(u,v)}{(a,w)} \, .
 \end{equation}
Thus, only one $\Psi$ can exist and it is given by the expression above. Consequently, we have obtained:
\begin{proposition} \label{prop-C2}
A non-geodesic time-like unit vector $u$ with $(w,a) \not=0$ is the velocity of a perfect energy tensor if, and only if,  the pair $\{u, \Psi\}$, where $\Psi=\Psi_2[u]$ is given in {\em (\ref{chi-C2})}, fulfills equations {\em (\ref{eq-uchi})}.

Then, the energy density $\rho$ and the pressure $p$ are determined as proposition {\em \ref{prop-anot=0-inverse}} states.
\end{proposition}
%


\subsection{Flows with $(w,a)=0$}
\label{subsec-(wa)=0}

In the following four sections we will study the cases with $(a,w)=0$. This condition and (\ref{chi-C2}) imply that $(u,v)=0$, and these two conditions are equivalent, respectively, to: 
\be \label{eq-(w,a)=0}
\dif u  \wedge a =0 \, , \qquad  \dif a  \wedge u \wedge a =0 \, .
\ee
The first condition means that a vector $b$ exists such that
\be \label{du-(w,a)=0}
\dif u = a \wedge u + a \wedge b \, , \quad (u,b)=0 \, , \quad (a,b)=0 \, .  \label{du}
\ee
Moreover, from here we can obtain $b$ as a concomitant of $u$:
\be \label{def-b}
\hspace{-8mm} b = b[u] \equiv - u +  \frac{1}{a^2} i(a) \dif u = -  \frac{1}{a^2}*(u \wedge a \wedge w)  \, , \quad w = *(u \wedge a \wedge b) \, .
\ee
Note that $w=0$ if, and only if, $b=0$. We will study separately the irrotational ($w=0$) and the rotating ($w \not=0$) flows. In these two cases, if we work with the equations (\ref{eq-uchi}) for the quantity $\Psi$ we obtain a tangled non-linear differential system. Thus, we prefer to study the initial conservation equations (\ref{con-eq1}-\ref{con-eq2}) and their integrability conditions, which become a linear system for $\{\rho, p\}$ and their derivatives. In this analysis, we must look for the equations that allow us to obtain the quotient $\dot{p}/(\rho+p) = \Psi$.

%
%
\section{Non-geodesic rotating flows: $a\not=0$, $w \not=0$, $(a,w)=0$. First results}
\label{sec-wnot=0_a}

Now, the expressions of the last subsection apply with $b \not=0$. Then, the tetrad $\{u, a, b, w\}$ defines an orthogonal basis. From now on, for a vector $x$, we write $x = x_u \, u + x_a \, a + x_b \, b + x_w w$,
\be \label{x-base}
x_u = - (x,u) \, , \quad   x_a = \frac{1}{a^2}(x,a) \, , \quad x_b = \frac{1}{b^2}(x,b) \, , \quad x_w = \frac{1}{w^2}(x,w) \, .
\ee
And for a function $\varphi$, 
\be \label{dphi-base}
\dot{\varphi} = -(\dif \varphi) _u \, , \quad \varphi^* = (\dif \varphi)_a \, , \quad \tilde{\varphi} = (\dif \varphi)_b  \, , \quad \hat{\varphi} = (\dif \varphi)_w  \, .
\ee

In this case, the integrability condition of (\ref{du}), $\dif^2u=0$, and (\ref{eq-(w,a)=0}) imply that vectors $c$ and $z$, and a function $\gamma$ exist such that
\begin{eqnarray}
\dif a = a \wedge c + \gamma \, b \wedge u, \qquad \quad (a, c)=0   \label{da} \, , \\[1mm]
\dif b = c \wedge (u+b) +  z \wedge a, \qquad (a, z)=0  \, . \label{db}
\end{eqnarray}
Moreover, $c$, $z$ and $\gamma$ can be obtained in terms of $u$ as:
\be \label{def-c}
\hspace{-20mm} c = c[u] \equiv  \frac{1}{a^2} i(a) \dif a    , \quad   z = z[u] \equiv - \frac{1}{a^2} i(a) \dif b    , \quad v = \gamma w , \quad \gamma = \gamma[u] \equiv \frac{(v,w)}{w^2}  .
\ee

The integrability condition of (\ref{da}), $\dif^2a=0$, implies that
\begin{equation} \label{dc}
\hspace{-10mm}  \dif c =(\gamma + \gamma^*) \ b \wedge  u + \gamma \, u \wedge z + y \wedge a  , \quad (a, y)=0 ,  \qquad \hat{\gamma} + 2 \gamma c_w =0    ,  \label{dc}
\end{equation}
where $y$ is a vector that can be obtained in terms of $u$ as:
\be  \label{def-y}
y = y[u] \equiv - \frac{1}{a^2} i(a) \dif c   \, .  
\ee

The conservation equations (\ref{con-eq1}-\ref{con-eq2}) can be written as:
\begin{equation}
\dif p = - \dot{p}\, u - \xi \, a  \, , \qquad 
\dot{\xi} = \dot{p} - \theta \xi \, ,  \qquad \xi \equiv \rho + p \, . \label{ce-eq_wnot=0}
\end{equation}
Then, a detailed analysis of the integrability conditions of these equations taking into account  (\ref{du-(w,a)=0}-\ref{da}-\ref{db}-\ref{dc}) leads to:
\begin{eqnarray}
\dif \dot{p} = \dot{p}\, \ell_1 + \xi \ell_2 + \rho^* \gamma u \, ,\label{dpdot} \\[1mm]
\dif \xi = \dot{p}\, \ell_3 + \xi \ell_4 + \rho^* a \, , \label{dxi}
\end{eqnarray}
where we have defined the following concomitants of $u$:
\begin{eqnarray}
\hspace{-10mm} \ell_1 \equiv \nu u - 2 a, \quad \ell_2 \equiv \pi u + \Omega a - \gamma b , \quad \ell_3 \equiv b-u, \quad \ell_4 \equiv \Omega u - a +c , \label{ell_i} \\
\hspace{-17mm} \pi = \pi[u] \equiv \gamma + \gamma^* + \tilde{\Omega} - \gamma z_b\, , \quad \nu = \nu[u]\equiv \Omega + 2\, c_b \, ,
\quad \Omega = \Omega[u] \equiv \theta  - c_u \, . \label{cpi_alpha}
\end{eqnarray}
We can now study the integrability conditions of equations (\ref{dpdot}-\ref{dxi}), $\dif^2 \dot{p}=0, \dif^2 \xi=0$, and we obtain:
\begin{eqnarray}
\hspace{-20mm}  \gamma\,  \dif \rho^*\! \wedge u =  \rho^* u \wedge e_1 + \dot{p} [ u \wedge e_2 +2 c_w\, a \wedge w] + \xi \, [u \wedge e_3 + (\hat{\Omega} \!-\! \gamma z_w) a \wedge w]  \label{ic_dpdot}  \, ,\\[1mm]
\hspace{-20mm} \dif \rho^{\!*} \! \wedge a =   \rho^* e_4 \wedge a + \dot{p}[ a \wedge e_5 +2 c_w\, u \wedge w]+ \xi \, [a \wedge e_6 + (\hat{\Omega} - \gamma z_w) u \wedge w]   \, . \label{ic_dxi}
\end{eqnarray}
where we have defined the following vectorial concomitants of $u$:
%
\begin{eqnarray}
\hspace{-20mm} e_1 \equiv \dif \gamma + (3 \gamma + \pi) a , \qquad \quad   e_2 \equiv \dif \nu +2 (\theta+ c_b\! - \! 2 c_u)  a+ (\pi\!- \!3 \gamma)  b  ,   \nonumber \\
\hspace{-20mm} e_3 \equiv \dif \pi + (2 \Omega  c_b \! +\! 2 \pi \! + \! \dot{\Omega}\! +\! \gamma z_u)  a  +[\gamma(\Omega \! - \! 2 c_b)\! -\! \dot{\gamma}\! +\! 2 \gamma c_u] b + (\gamma^* \! \!+\! \tilde{\Omega} \!- \! \gamma z_b) c  ,  \label{e_i}  \\ 
\hspace{-20mm} e_4 \equiv \Omega \, u + 2 c \, , \qquad  \quad e_5 \equiv 2 b + z   ,  \qquad \quad   e_6 \equiv c  +y -\Omega^*  u\! - \!2 \Omega\, b     . \nonumber
\end{eqnarray}
%
And, if we calculate the $u\wedge \omega$ component of (\ref{ic_dxi}) we obtain:
\begin{equation}
2\, \dot{p} \, c_w + \xi (\hat{\Omega} - \gamma z_w) =0  \, . \label{c-w_a}
\end{equation}
%

%
%
\subsection{Case $(c,w)\not=0$ {\em (C$_3$)}}
\label{subsec-wnot=0;(cw)not=0}

Now, from (\ref{c-w_a}) we can determine $\Psi = \dot{p}/\xi$ as:
\begin{equation} \label{chi-(cw)not=0}
\Psi = \Psi_3[u] \equiv  \frac{\gamma z_w - \hat{\Omega}}{2 \, c_w}  \, .
 \end{equation}
This expression determines $\Psi$ in terms of $u$ and, consequently, proposition \ref{prop-anot=0} applies:
\begin{proposition} \label{prop-(cw)not=0}
A non-geodesic time-like unit vector $u$ with $(a,w)=0$, and $(c,w)\not=0$, is the velocity of a perfect energy tensor if, and only if,  the pair $\{u, \Psi\}$, where $\Psi=\Psi_3[u]$ is given in {\em (\ref{chi-(cw)not=0})}, fulfills equations {\em (\ref{eq-uchi})}.

Then, the energy density $\rho$ and the pressure $p$ are determined as proposition {\em \ref{prop-anot=0-inverse}} states.
\end{proposition}
%

%
\subsection{Some constraints for the case $(c,w)=0$}
\label{subsec-wnot=0;(cu)=0}

When $(c,w)=0$, (\ref{c-w_a}) and (\ref{dc}) imply, respectively, $\hat{\Omega} - \gamma z_w=0$ and $\hat{\gamma}=0$. Then, by putting $c=c_u u + c_b b$ and equaling its differential with (31) we obtain $\hat{c}_b=0$ and then $\hat{\nu}=\hat{\Omega}$. Taking the differential of $\pi$ from (37) we also get $\hat{\pi}-\gamma y_w = 0$. To obtain this, we have used $\dif ^2 c = 0$ from (31) and the resulting condition after taking into account that $\dif \Omega - \gamma z$ is orthogonal to $w$ and differentiating it. Thus, we have the following constraints:
\begin{equation}\label{constraints}
 c_w = \hat{\gamma} = \hat{c_b}=  \hat{\Omega} - \gamma z_w = \hat{\nu} - \gamma z_w = \hat{\pi} - \gamma y_w =0   \, .
\end{equation}
Note that the integrability conditions (\ref{ic_dpdot}-\ref{ic_dxi}) fully determine $\dif \rho^*$ when $\gamma \not=0$, and only partially when $\gamma=0$. Thus, we will consider these two cases separately, which correspond, respectively, to $v \not=0$ and $v=0$.

%
%
\section{Non-geodesic rotating flows: $a \not=0 \not=w$, $(a,w)\!=\!0$. Case $(c,w)\!=\!0$, $v\!\not=\!0$}
\label{sec-wnot=0_b}

Now, $\gamma \not=0$ as a consequence of (\ref{def-c}). Then, (\ref{ic_dpdot}-\ref{ic_dxi}) and the constraints (\ref{constraints}) imply:
\begin{eqnarray}
\dif \rho^* = \dot{p}\, f_1 + \xi \, f_2 +  \rho^* f_3 \, , \label{drho*}   \\[1mm]
 \dot{p}\, \, \phi_1 + \xi \,  \phi_2 +  \rho^* \phi_3 =0    \, , \label{algebraica}
\end{eqnarray}
where the vectors $f_i$ and the scalars $\phi_i$ are concomitants of $u$ given by
\begin{eqnarray}
f_1 \equiv -\gamma^{-1}[\nu^* +2 (\theta+ c_b\! - \! 2 c_u)] \, a - 2  b -z \, ,   \nonumber \\
f_2 \equiv\Omega^* u -  \gamma^{-1}[\pi^* \!+ \! \dot{\Omega} + \gamma z_u +2 (\Omega c_b\! + \! \pi)] \, a+ 2 \Omega \, b \! - c - y \, , \label{f_i} \\ 
f_3 \equiv \Omega  u\! - \gamma^{-1}[\gamma^* + 3 \gamma + \pi] \, a  + 2c     \, , \nonumber
\end{eqnarray}
\begin{eqnarray}
\phi_1 = \phi_1[u] \equiv \tilde{\nu} - 5 \gamma + \pi - \gamma z_b \, ,  \nonumber \\
\phi_2 = \phi_2[u]  \equiv \tilde{\pi} - \dot{\gamma} - \gamma(c_u\! +\! 4 c_b \! -\!3 \theta \! + \! y_b) + \pi c_b  \, ,  \label{phi_i} \\
\phi_3 = \phi_3[u]  \equiv \tilde{\gamma} + 2 \gamma c_b  \,   . \nonumber
\end{eqnarray}
Then, equations (\ref{dpdot}-\ref{dxi}-\ref{drho*}) define an exterior differential system for the functions $\{\dot{p}, \xi, \rho^*\}$, with an additional algebraic constraint given by (\ref{algebraica}). Note that the integrability conditions of equations (\ref{dpdot}-\ref{dxi}) are equivalent to (\ref{drho*}) and (\ref{algebraica}). Thus, when studying the differential system, we only need to consider the integrability conditions of (\ref{drho*}), and as a consequence of (\ref{ic_dpdot}-\ref{ic_dxi}), only the component $u\! \wedge\! a$ of $\dif^2 \rho^*$.

%
%
\subsection{Case $\phi_3\not=0$}
\label{subsec-phi_3not=0}

Now, from the algebraic equation (\ref{algebraica}) we can obtain: 
\be
\rho^* \!= - \bar{\phi}_1 \dot{p} -  \bar{\phi}_2 \xi \, , \qquad  \bar{\phi}_1 \equiv \frac{\phi_1}{ \phi_3}  \, , \quad  \bar{\phi}_2 \equiv \frac{\phi_2}{ \phi_3}  \, . \label{rho*}
\ee
Then, substituting this expression and its differential in (\ref{drho*}) we get:
\begin{equation}
 \dot{p}\, \, h_1 + \xi \,  h_2 =0   \, ,  \label{eq-h_i} 
\end{equation}
$h_i$ being the vectorial concomitants of $u$ given by
\begin{equation}
\begin{array}{rl}
\hspace{-10mm}   h_1 = h_1[u]  \equiv \dif \bar{\phi}_1 +  \bar{\phi}_1[\ell_1 -  \bar{\phi}_1 \gamma u]+  \bar{\phi}_2 [\ell_3 -  \bar{\phi}_1 a] + f_1 -  \bar{\phi}_1 f_3 \, ,  \label{h_1}  \\[1mm]
\hspace{-10mm}   h_2 = h_2[u]  \equiv \dif \bar{\phi}_2 +  \bar{\phi}_1 [\ell_2 -  \bar{\phi}_2 \gamma u]+   \bar{\phi}_2[\ell_4 - \bar{\phi}_2 a] + f_2 -  \bar{\phi}_2 f_3  \label{h_2}  \, , 
\end{array}
\end{equation}
where the vector concomitants $\ell_i$ and $f_i$ are defined in (\ref{ell_i}) and (\ref{f_i}), respectively.
%

%
%
\subsubsection{Subcase $h_1 \not= 0$ {\em (C$_4$)}}
\label{subsubsec-phi_3not=0-1}

\ \\[2mm]
Now, from (\ref{eq-h_i}) we can determine $\Psi = \dot{p}/\xi$ as:
\begin{equation} 
\Psi = \Psi_4[u] \equiv -  \frac{\bar{g}(h_1, h_2)}{\bar{g}(h_1, h_1)}  \, , \qquad \bar{g} = \bar{g}[u] \equiv g + 2 u \otimes u \, . \label{g_tilde}
 \end{equation}
 Note that $\bar{g}$ is a Riemannian metric and, thus, $\bar{g}(h_1, h_1)>0$. Expression (\ref{g_tilde}) determines $\Psi$ in terms of $u$ and, consequently, proposition \ref{prop-anot=0} applies:
\begin{proposition} \label{prop-h_1not=0}
A non-geodesic time-like unit vector $u$ with $w \not=0$, $(a,w)=(c,w)=0$, $v\not=0$, $\phi_3 \not=0$, and $h_1 \not=0$, is the velocity of a perfect energy tensor if, and only if, the pair $\{u, \Psi\}$, where $\Psi=\Psi_4[u]$ is given in {\em (\ref{g_tilde})}, fulfills equations {\em (\ref{eq-uchi})}.

Then, the energy density $\rho$ and the pressure $p$ are determined as proposition {\em \ref{prop-anot=0-inverse}} states.
\end{proposition}
%

%
%
\subsubsection{Subcase $h_1=0$ {\em (C$_5$)}} 
\label{subsubsec-phi_3not=0-2}

\ \\[2mm]
Now, equation (\ref{eq-h_i}) implies that, necessarily, $h_2=0$. Moreover, if we use the expression (\ref{rho*}) for $\rho^*$ and substitute it in (\ref{dpdot}-\ref{dxi}), we obtain the following exterior system for the functions $\{\dot{p}, \xi\}$:
\begin{equation}
\dif \dot{p} = \dot{p}\, g_1 + \xi \, g_2  \, , \qquad \dif \xi = \dot{p}\, g_3 + \xi \, g_4  \, ,  \label{dpdot_dxi} 
\end{equation}
\begin{equation}
\hspace{-15mm} g_1 \equiv \ell_1- \gamma \bar{\phi}_1 u  , \quad \ g_2 \equiv \ell_2 - \gamma \bar{\phi}_2 u   , \quad \
g_3 \equiv  \ell_3 - \bar{\phi}_1 a   , \quad  \ g_4 \equiv \ell_4 -  \bar{\phi}_2 a    ,\label{g_i}  
\end{equation}
where the vector concomitants $\ell_i$ are defined in (\ref{ell_i}).

The integrability conditions of the exterior system (\ref{dpdot_dxi}) are equivalent to (\ref{eq-h_i}), which identically holds when $h_1=h_2=0$. Thus, the system (\ref{dpdot_dxi}) admits solution. Moreover, equations (\ref{ce-eq_wnot=0}) determine the pair $\{\rho, p\}$ that completes the solution of the conservation equations. We summarize this result as follows.
\begin{proposition} \label{prop-h_1=0}
A non-geodesic time-like unit vector $u$ with $w \not=0$, $(a,w)=(c,w)=0$, $v\not=0$, $\phi_3 \not=0$, and $h_1=0$, is the velocity of a perfect energy tensor if, and only if, $(u,v)=0$, $\hat{\Omega} = \gamma z_w$ and $h_2=0$.

Then, the energy density $\rho$ and the pressure $p$ are determined by {\em (\ref{ce-eq_wnot=0})}, where $\{\dot{p}, \xi\}$ is the general solution of the exterior system {\em (\ref {dpdot_dxi})}.
\end{proposition}
The integration method of the exterior system (\ref{dpdot_dxi}) and the analysis of its richness of solutions will be analyzed in \ref{ap-A}.

%
%
\subsection{Case $\phi_3=0$, $\phi_1 \not=0$ {\em (C$_6$)}}
\label{subsec-phi_3=0-phi_2not=0}

Under these constraints, equation (\ref{algebraica}) enables us to determine $\Psi = \dot{p}/\xi$ as:
\begin{equation} \label{chi-6}
\Psi = \Psi_6[u] \equiv  - \frac{\phi_2}{\phi_1}  \, .
 \end{equation}
This expression determines $\Psi$ in terms of $u$ and, consequently, proposition \ref{prop-anot=0} applies:
\begin{proposition} \label{prop-(cw)not=0}
A non-geodesic time-like unit vector $u$ with $w \not=0$, $(a,w)=(c,w)=0$, $v\not=0$, $\phi_3=0$, and $\phi_1 \not=0$, is the velocity of a perfect energy tensor if, and only if, the pair $\{u, \Psi\}$, where $\Psi=\Psi_6[u]$ is given in {\em (\ref{chi-6})}, fulfills equations {\em (\ref{eq-uchi})}.

Then, the energy density $\rho$ and the pressure $p$ are determined as proposition {\em \ref{prop-anot=0-inverse}} states.
\end{proposition}
%

%

\subsection{Case $\phi_3=0$, $\phi_1 =0$}
\label{subsec-phi_3=0-phi_2=0}

Now, equation (\ref{algebraica}) implies that, necessarily, $\phi_2=0$. Then, the algebraic constraint (\ref{algebraica}) is an identity, and the sole integrability condition that we must impose on the exterior system (\ref{dpdot}-\ref{dxi}-\ref{drho*}) is the component $u\! \wedge \!a$ of $\dif^2 \rho^*=0$, which leads to:
\begin{equation}
 \dot{p}\, \, \kappa_1 + \xi \,  \kappa_2 +  \rho^* \kappa_3 =0    \, , \qquad  \kappa_i  =\kappa_i[u] \equiv i(u)i(a)K_i \, .\label{ic_kappa_i} 
\end{equation}
$K_i$ being the 2-forms given by
\begin{eqnarray}
K_1 \equiv \dif f_1 + \ell_1 \wedge f_1 + \ell_3 \wedge f_2 + f_1 \wedge f_3 \, ,  \nonumber \\
K_2 \equiv \dif f_2 + \ell_2 \wedge f_1 + \ell_4 \wedge f_2 + f_2 \wedge f_3  \, ,  \label{K_i} \\
K_3 \equiv \dif f_3 + \gamma u \wedge f_1 + a \wedge f_2   \,   , \nonumber
\end{eqnarray}
where $e_i$ and $f_i$ are defined in (\ref{e_i}) and (\ref{f_i}), respectively. 
%

%

\subsubsection{Case $\kappa_3\not=0$}
\label{subsubsec-kappa_3not=0}

\ \\[2mm]
Now, from the algebraic equation (\ref{ic_kappa_i}) we can obtain: 
\be
\rho^* \!= - \bar{\kappa}_1 \dot{p} - \bar{\kappa}_2 \xi \, , \qquad  \bar{\kappa}_1 \equiv \frac{\kappa_1}{ \kappa_3} \, , \quad  \bar{\kappa}_2 \equiv \frac{\kappa_2}{ \kappa_3} \, . \label{rho*kappa_i}
\ee
Then, substituting this expression and its differential in (\ref{drho*}) we obtain:
\begin{equation}
 \dot{p}\, \, q_1 + \xi \,  q_2 =0   \, ,  \label{eq-q_i} 
\end{equation}
$q_i$ being the vectorial concomitants of $u$ given by
\begin{equation}
\begin{array}{l}
\hspace{-10mm}   q_1 = q_1[u] \equiv \dif \bar{\kappa}_1 + \bar{\kappa}_1[\ell_1 - \bar{\kappa}_1 \gamma u]  +  \bar{\kappa}_2 [\ell_3 - \bar{\kappa}_1 a] + f_1 - \bar{\kappa}_1 f_3 \, , \label{q_1}   \\[1mm]
\hspace{-10mm}   q_2 = q_2[u]  \equiv \dif  \bar{\kappa}_2+ \bar{\kappa}_1[\ell_2 - \bar{\kappa}_2 \gamma u]+  \bar{\kappa}_2 [\ell_4 -\bar{\kappa}_2 a] + f_2 - \bar{\kappa}_2 f_3  \label{q_2}  \,  , 
\end{array}
\end{equation}
where $\ell_i$ and $f_i$ are defined in (\ref{ell_i}) and (\ref{f_i}), respectively.
%
%
%
\ \\[2mm]
\noindent
{\em A. Subcase $\kappa_3 \not= 0$, $q_1 \not=0$ {\em (C$_7$)}}
\ \\[2mm]
Now, from (\ref{eq-q_i}) we can determine $\Psi = \dot{p}/\xi$ as:
\begin{equation} 
\Psi = \Psi_7[u] \equiv -  \frac{\bar{g}(q_1, q_2)}{\bar{g}(q_1, q_1)}  \, , \qquad \bar{g} = \bar{g}[u] \equiv g + 2 u \otimes u \, . \label{g_tilde_q_i}
 \end{equation}
 Note that $\bar{g}$ is a Riemannian metric and, thus, $\bar{g}(q_1, q_1)>0$. Expression (\ref{g_tilde_q_i}) determines $\Psi$ in terms of $u$ and, consequently, proposition \ref{prop-anot=0} applies:
\begin{proposition} \label{prop-h_1not=0}
A non-geodesic time-like unit vector $u$ with $w \not=0$, $(a,w)=(c,w)=0$, $v\not=0$, $\phi_3 = \phi_1=0$, $\kappa_3 \not=0$, and $q_1\not=0$, is the velocity of a perfect energy tensor if, and only if, the pair $\{u, \Psi\}$, where $\Psi=\Psi_7[u]$ is given in {\em (\ref{g_tilde_q_i})}, fulfills equations {\em (\ref{eq-uchi})}.

Then, the energy density $\rho$ and the pressure $p$ are determined as proposition {\em \ref{prop-anot=0-inverse}} states.
\end{proposition}
%
%
%
%
%
\ \\[0mm]
\noindent
{\em B. Subcase $\kappa_3 \not= 0$, $q_1 =0$ {\em (C$_8$)}}
\ \\[2mm]
Now, equation (\ref{eq-q_i}) implies that, necessarily, $q_2=0$. Moreover, if we use the expression (\ref{rho*kappa_i}) for $\rho^*$ and substitute it in (\ref{dpdot}-\ref{dxi}), we obtain the following exterior system for the functions $\{\dot{p}, \xi\}$:
\begin{equation}
\dif \dot{p} = \dot{p}\, m_1 + \xi \, m_2  \, , \qquad \dif \xi = \dot{p}\, m_3 + \xi \, m_4  \, ,  \label{dpdot_dxi_n_i} 
\end{equation}
\begin{equation}
\hspace{-15mm} m_1 \equiv \ell_1- \gamma \bar{\kappa}_1 u  , \quad \ m_2 \equiv \ell_2 - \gamma \bar{\kappa}_2 u   , \quad \
m_3 \equiv  \ell_3 - \bar{\kappa}_1 a   , \quad  \ m_4 \equiv \ell_4 -  \bar{\kappa}_2 a    ,\label{m_i}  
\end{equation}
where the vector concomitants $\ell_i$ are defined in (\ref{ell_i}).

The integrability conditions of the exterior system (\ref{dpdot_dxi_n_i}) are equivalent to (\ref{eq-q_i}), which identically holds when $q_1=q_2=0$. Thus, the system (\ref{dpdot_dxi_n_i}) admits solution. Moreover, equations (\ref{ce-eq_wnot=0}) determine the pair $\{\rho, p\}$ that completes the solution of the conservation equations. We summarize this result below.
\begin{proposition} \label{prop-q_1=0}
A non-geodesic time-like unit vector $u$ with $w \not=0$, $(a,w)=(c,w)=0$, $v\not=0$, $\phi_3 = \phi_1=0$, $\kappa_3 \not=0$, and $q_1=0$, is the velocity of a perfect energy tensor if, and only if, $(u,v)=0$, $\hat{\Omega} = \gamma z_w$, $\phi_2=0$ and $q_2=0$.

Then, the energy density $\rho$ and the pressure $p$ are determined by {\em (\ref{ce-eq_wnot=0})}, where $\{\dot{p}, \xi\}$ is the general solution of the exterior system {\em (\ref {dpdot_dxi_n_i})}.
\end{proposition}
The integration method of the exterior system (\ref{dpdot_dxi_n_i}) and the analysis of its richness of solutions will be analyzed in \ref{ap-A}.

%
%
\subsubsection{Case $\kappa_3=0$, $\kappa_1 \not=0$ {\em (C$_9$)}}
\label{subsubsec-kappa_3=0-kappa_2not=0}

\ \\[2mm]
Under these constraints, equation (\ref{ic_kappa_i}) enables us to determine $\Psi = \dot{p}/\xi$ as:
\begin{equation} \label{kappa_1not=0}
\Psi = \Psi_9[u] \equiv  - \frac{\kappa_2}{\kappa_1}  \, .
 \end{equation}
This expression determines $\Psi$ in terms of $u$ and, consequently, proposition \ref{prop-anot=0} applies:
\begin{proposition} \label{prop-kappa_1not=0}
A non-geodesic time-like unit vector $u$ with $w \not=0$, $(a,w)=(c,w)=0$, $v\not=0$, $\phi_3 = \phi_1=0$, $\kappa_3 = 0$, and $\kappa_1 \not=0$, is the velocity of a perfect energy tensor if, and only if, the pair $\{u, \Psi\}$, where $\Psi=\Psi_9[u]$ is given in {\em (\ref{kappa_1not=0})}, fulfills equations {\em (\ref{eq-uchi})}.

Then, the energy density $\rho$ and the pressure $p$ are determined as proposition {\em \ref{prop-anot=0-inverse}} states.
\end{proposition}
%

%
%
\subsubsection{Case $\kappa_3=0$, $\kappa_1 =0$ {\em (C$_{10}$)}}
\label{subsubsubsec-kappa_3=0-phi_2not=0}

\ \\[2mm]
In this case, equation (\ref{ic_kappa_i}) implies that, necessarily, $\kappa_2=0$. Then, the integrability conditions of the exterior system (\ref{dpdot}-\ref{dxi}-\ref{drho*}) identically hold and, consequently, it admits solution $\{\dot{p}, \xi, \rho^*\}$. Moreover, equations (\ref{ce-eq_wnot=0}) determine the pair $\{\rho, p\}$ that completes the solution of the conservation equations. We summarize this result as follows.
\begin{proposition} \label{prop-kappa_i=0}
A non-geodesic time-like unit vector $u$ with $w \not=0$, $(a,w)=(c,w)=0$, $v\not=0$, $\phi_3 = \phi_1=0$, and $\kappa_3 = \kappa_1=0$, is the velocity of a perfect energy tensor if, and only if, $(u,v)=0$, $\hat{\Omega} = \gamma z_w$, $\phi_2=0$ and $\kappa_2=0$.

Then, the energy density $\rho$ and the pressure $p$ are determined by {\em (\ref{ce-eq_wnot=0})}, where $\{\dot{p}, \xi, \rho^*\}$ is the general solution of the exterior system {\em (\ref{dpdot}-\ref{dxi}-\ref{drho*})}.
\end{proposition}
The integration method of the exterior system (\ref{dpdot}-\ref{dxi}-\ref{drho*}) and the analysis of its richness of solutions will be analyzed in \ref{ap-B}.

%

%
%
\section{Non-geodesic rotating flows: $a \not=0 \not=w$, $(a,w)\!=\!0$. Case $(c,w)\!=\!0$, $v\!=\!0$}
\label{sec-wnot=0_c}

Now $\gamma=0$, and we must consider the exterior system (\ref{dpdot}-\ref{dxi}), its integrability conditions (\ref{ic_dpdot}-\ref{ic_dxi}), and constraints (\ref{constraints}), for this case. We have that (\ref{ic_dpdot}) becomes an algebraic constraint for $\{\dot{p}, \xi, \rho^*\}$, which is equivalent to its components $u\! \wedge \!a$ and $u\! \wedge \!b$:
\begin{eqnarray}
2\, \dot{p} \, [\tilde{c_b} + \tilde{\Omega}] + \xi \,  [\tilde{\tilde{\Omega}} + \tilde{\Omega} c_b]  =0 \, ,   \qquad \dot{p} \, \varphi_1 + \xi \,  \varphi_2+  \rho^* \tilde{\Omega} =0\, , \label{gamma=0_ic-a} \\[2mm]
\varphi_1 \equiv \nu^* \!+ 2(\theta + c_b - 2c_u)  \, , \qquad \varphi_2 \equiv \tilde{\Omega}^* \! + \dot{\Omega}+ 2(\Omega c_b + \tilde{\Omega})  \, .
  \label{varphi_i}
\end{eqnarray}
%
%
%

\subsection{Case $\tilde{c_b} + \tilde{\Omega} \not=0$ {\em (C$_{11}$)}}
\label{subsec-hat_not=0}

Now, from the first equation in (\ref{gamma=0_ic-a}) we can determine $\Psi = \dot{p}/\xi$ as:
\begin{equation} 
 \Psi = \Psi_{11}[u] \equiv -  \frac{\tilde{\tilde{\Omega}} + \tilde{\Omega}\, c_b}{2(\tilde{c_b} + \tilde{\Omega})} \, .\label{Psi_alpha_tilde}
 \end{equation}
 Expression (\ref{Psi_alpha_tilde}) determines $\Psi$ in terms of $u$ and, consequently, proposition \ref{prop-anot=0} applies:
\begin{proposition} \label{prop-h_1not=0}
A non-geodesic time-like unit vector $u$ with $w \not=0$, $(a,w)=(c,w)=0$, $v=0$, and $\hat{\Omega} + \tilde{\Omega} \not=0$, is the velocity of a perfect energy tensor if, and only if, the pair $\{u, \Psi\}$, where $\Psi=\Psi_{11}[u]$ is given in {\em (\ref{Psi_alpha_tilde})}, fulfills equations {\em (\ref{eq-uchi})}.

Then, the energy density $\rho$ and the pressure $p$ are determined as proposition {\em \ref{prop-anot=0-inverse}} states.
\end{proposition}
%

%
%

\subsection{Case $\tilde{c_b} =\! - \tilde{\Omega}\not=0$}
\label{subsec-hat=0_tildenot=0}

Now, equation (\ref{gamma=0_ic-a}) implies that, necessarily, $\tilde{\tilde{\Omega}} + \tilde{\Omega}\, c_b =0$. Moreover, from the second algebraic equation in (\ref{gamma=0_ic-a}) we can obtain: 
\be
\rho^* \!= -\bar{\varphi}_1 \dot{p} - \bar{\varphi}_2 \xi \, , \qquad \bar{\varphi}_1 \equiv \frac{\varphi_1}{\tilde{\Omega}} \, , \qquad   \bar{\varphi}_2 \equiv \frac{\varphi_2}{\tilde{\Omega}}  \, . \label{rho*_gamma=0}
\ee
Then, substituting this expression and its differential in (\ref{ic_dxi}), and using constraints (\ref{constraints}), we obtain an equation of the form $n \wedge a=0$, which is equivalent to $n \!- \!n_a a\!=\!0$. This vectorial equation can be written as
\begin{equation}
 \dot{p}\, \, n_1 + \xi \,  n_2 =0   \, ,  \label{eq-n_i} 
\end{equation}
$n_i$ being the vectorial concomitants of $u$ given by
\begin{equation}
\begin{array}{l}
\hspace{-20mm} n_1 = n_1[u] \equiv \dif \bar{\varphi}_1 +\bar{\varphi}_2 \ell_3   + \bar{\varphi}_1 \, (\ell_1 - \bar{\varphi}_1 \gamma u)+ (2 \bar{\varphi}_1 - \bar{\varphi}_1^*)a -  \bar{\varphi}_1  e_4 -e_5 \, ,  \label{n_1} \\[1mm]
\hspace{-20mm} n_2 = n_2[u] \equiv \dif \bar{\varphi}_2 +\bar{\varphi}_1 \ell_2   + \bar{\varphi}_2 \, (\ell_4 - \bar{\varphi}_1 \gamma u)+ (\bar{\varphi}_2 - \Omega \bar{\varphi}_1 - \bar{\varphi}_2^*)a -  \bar{\varphi}_2  e_4 -e_6  \, , \label{n_2}
\end{array}
\end{equation}
where $e_i$ and $\ell_i$ are defined in (\ref{e_i}) and (\ref{ell_i}), respectively.
%

%
%
\subsubsection{Subcase $n_1 \not= 0$ {\em (C$_{12}$)}}
\label{subsubsec-n_1not=0}

\ \\[2mm]
Now, from (\ref{eq-n_i}) we can determine $\Psi = \dot{p}/\xi$ as:
\begin{equation} 
\Psi = \Psi_{12}[u] \equiv -  \frac{\bar{g}(n_1, n_2)}{\bar{g}(n_1, n_1)}  \, , \qquad \bar{g} = \bar{g}[u] \equiv g + 2 u \otimes u \, . \label{g_bar_3}
 \end{equation}
 Note that $\bar{g}$ is a Riemannian metric and, thus, $\bar{g}(n_1, n_1)\!>\!0$. Expression (\ref{g_bar_3}) determines $\Psi$ in terms of $u$ and, consequently, proposition \ref{prop-anot=0} applies:
\begin{proposition} \label{prop-h_1not=0}
A non-geodesic time-like unit vector $u$ with $w \not=0$, $(a,w)=(c,w)=0$, $v=0$, $\tilde{c_b} =\! - \tilde{\Omega}\not=0$, and $n_1 \not=0$, is the velocity of a perfect energy tensor if, and only if, the pair $\{u, \Psi\}$, where $\Psi=\Psi_{12}[u]$ is given in {\em (\ref{g_bar_3})}, fulfills equations {\em (\ref{eq-uchi})}.

Then, the energy density $\rho$ and the pressure $p$ are determined as proposition {\em \ref{prop-anot=0-inverse}} states.
\end{proposition}
%

%
%
\subsubsection{Subcase $n_1=0$ {\em (C$_{13}$)}} 
\label{subsubsec-n_1=0}

\ \\[2mm]
Now, equation (\ref{eq-n_i}) implies that, necessarily, $n_2=0$. Moreover, if we use the expression (\ref{rho*_gamma=0}) for $\rho^*$ and substitute it in (\ref{dpdot}-\ref{dxi}), we obtain the following exterior system for the functions $\{\dot{p}, \xi\}$:
\begin{equation}
\dif \dot{p} = \dot{p}\, \bar{\ell}_1 + \xi \, \bar{\ell}_2  \, , \qquad \dif \xi = \dot{p}\, \bar{\ell}_3 + \xi \, \bar{\ell}_4  \, ,  \label{dpdot_dxi_gamma=0} 
\end{equation}
\begin{equation}
\bar{\ell}_1 \equiv  \ell_1\, , \quad \bar{\ell}_2 \equiv  \ell_2 \, , \quad     \bar{\ell}_3 \equiv  \ell_3 - \bar{\varphi}_1 a  \, , \quad   \bar{\ell}_4 \equiv \ell_4 -   \bar{\varphi}_2 a   \, ,\label{barell_i}  
\end{equation}
where the vector concomitants $\ell_i$ are defined in (\ref{ell_i}).

The integrability conditions of the exterior system (\ref{dpdot_dxi_gamma=0}) are equivalent to (\ref{eq-n_i}), which identically holds when $n_1=n_2=0$. Thus, the system (\ref{dpdot_dxi_gamma=0}) admits solution. Moreover, equations (\ref{ce-eq_wnot=0}) determine the pair $\{\rho, p\}$ that completes the solution of the conservation equations. We summarize this result as follows.
\begin{proposition} \label{prop-h_1=0}
A non-geodesic time-like unit vector $u$ with $w \not=0$, $(a,w)=(c,w)=0$, $v=0$, $\tilde{c_b} =\! - \tilde{\Omega}\not=0$, and $n_1=0$, is the velocity of a perfect energy tensor if, and only if, $\hat{\Omega}=0$, $\tilde{\tilde{\Omega}} + \tilde{\Omega}\, c_b =0$ and $n_2=0$.

Then, the energy density $\rho$ and the pressure $p$ are determined by {\em (\ref{ce-eq_wnot=0})}, where $\{\dot{p}, \xi\}$ is the general solution of the exterior system {\em (\ref {dpdot_dxi_gamma=0})}.
\end{proposition}
The integration method of the exterior system (\ref{dpdot_dxi_gamma=0}) and the analysis of its richness of solutions will be analyzed in \ref{ap-A}.

%
%
\subsection{Case $\tilde{c_b} = \tilde{\Omega}=0$, $\varphi_1 \not=0$ {\em (C$_{14}$)}}
\label{subsec-hat=0_tildenot=0-A}

When $\tilde{c_b} = \tilde{\Omega}=0$, we have $\varphi_2 =2 \Omega c_b + \dot{\Omega}$, and the second equation in (\ref{gamma=0_ic-a}) enables us to determine $\Psi = \dot{p}/\xi$ as:
\begin{equation} \label{Omegues=0}
\Psi = \Psi_{14}[u] \equiv  - \frac{2 \Omega c_b + \dot{\Omega}}{\varphi_1}  \, .
 \end{equation}
This expression determines $\Psi$ in terms of $u$ and, consequently, proposition \ref{prop-anot=0} applies:
\begin{proposition} \label{prop-(cw)not=0}
A non-geodesic time-like unit vector $u$ with $w \not=0$, $(a,w)=(c,w)=0$, $v=0$, $\tilde{c_b} =\!  \tilde{\Omega}=0$, and $\varphi_1 \not=0$, is the velocity of a perfect energy tensor if, and only if,  the pair $\{u, \Psi\}$, where $\Psi=\Psi_{14}[u]$ is given in {\em (\ref{Omegues=0})}, fulfills equations {\em (\ref{eq-uchi})}.

Then, the energy density $\rho$ and the pressure $p$ are determined as proposition {\em \ref{prop-anot=0-inverse}} states.
\end{proposition}
%

%

\subsection{Case $\tilde{c_b} = \tilde{\Omega}=0$, $\varphi_1=0$ {\em (C$_{15}$)}}
\label{subsec-hat=0_tildenot=0_B}

In this case, equation (\ref{gamma=0_ic-a}) leads to $\varphi_2 =2 \Omega c_b + \dot{\Omega}=0$. Then, the  integrability conditions identically hold and we can find the solution of the conservation equations as follows. The constraints imposed on $u$ in this class imply that three independent functions $\{\tau, \alpha, \mu \}$ and a function $\beta = \beta(\tau, \alpha, \mu)$ exist such that:
\begin{equation} \label{tau,alpha,lambda}
u=-\dif \tau + \mu \, \dif \alpha \, , \qquad a =\beta \, \dif \alpha  \, .
 \end{equation}
Then, the conservation equations (\ref{con-eq1}-\ref{con-eq2}) imply $p=p(\tau,\alpha)$, and they are equivalent to:
 \begin{eqnarray} \label{rho-p-tau,alpha}
\beta(\rho + p) +\mu \, p_{,\tau } + p_{,\alpha}=0 \, , \\ 
p_{,\tau \tau} +\nu p_{,\tau} =0 \,  , \qquad p_{,\tau \alpha} + \Omega p_{,\alpha} + 2 (\beta- \mu c_b) \,p_{,\tau} =0   \,  .  \label{p-tau,alpha}
 \end{eqnarray}
Equations (\ref{p-tau,alpha}) define a second-order partial differential system for the function $p(\tau, \alpha)$, which always admits solution because its integrability conditions identically hold. Moreover, for every solution $p=p(\tau, \alpha)$ to this system, equation (\ref{rho-p-tau,alpha}) determines $\rho$. Consequently, we obtain:
\begin{proposition} \label{prop-kappa_i=0}
A non-geodesic time-like unit vector $u$ with $w \not=0$, $(a,w)=(c,w)=0$, $v=0$, $\tilde{c_b} =\!  \tilde{\Omega}=0$, and $\varphi_1=0$, is the velocity of a perfect energy tensor if, and only if, $\hat{\Omega} =0$ and $2 \Omega c_b + \dot{\Omega}=0$.

Then, three independent functions $\{\tau, \alpha, \mu \}$ and a function $\beta = \beta(\tau, \alpha, \mu)$ exist such that $u\!=\!-\dif \tau \!+\! \mu \, \dif \alpha$, $a\! =\! \beta  \, \dif \alpha$, and the energy density $\rho$ and the pressure $p$ are given by:
\begin{equation}
\rho = - p - \beta^{-1}[\mu\, p_{,\tau } + p_{,\alpha}]  \, , \qquad p=p(\tau, \alpha) \, ,  \label{rhode-p-C}
\end{equation}
where $p(\tau, \alpha)$ is any solution of {\em (\ref{p-tau,alpha})}.
\end{proposition}
%

%
%
\section{Non-geodesic irrotational flows: $a\not=0$, $w =0$}
\label{sec-w=0}

Now, the vector $b$ defined in (\ref{def-b}) vanishes. Then, equations (\ref{eq-(w,a)=0}) imply that a vector $m$ exists such that:
\begin{equation} \label{duda-w=0}
\dif u = a \wedge u, \qquad \dif a= u \wedge m \, , \qquad (m,u)=0 \, ,
\end{equation}
%
%
%
Moreover, $m$ can be obtained in terms of $u$ as:
\be \label{def-m}
m = m[u] \equiv - i(u) \dif a \, .
\ee
Additionally, the following property holds:
\be \label{u-a-m-v}
\hspace{-20mm} v = *(a \wedge \dif a) =*(u \wedge m \wedge a) \not = 0  \quad \leftrightarrow \quad  m \wedge a \not =0 \quad \leftrightarrow \quad \{u,a,v,m\} \ \ {\rm basis}  \, .
\ee

Equations (\ref{duda-w=0}) and the integrability condition of the conservation equation (\ref{con-eq1}), $\dif^2 p =0$, implies $\dif \xi \wedge u \wedge a = 0$, where $\xi = \rho + p$. In this case, the conservation equations (\ref{con-eq1}-\ref{con-eq2}) become:
\begin{eqnarray}
\dif p = - \dot{p}\, u - \xi \, a \label{ce-eq1_0} \, ,\\
\dif \xi = (\theta \xi - \dot{p}) \, u + (\rho^{\!*} - \xi) \, a \, . \label{ce-eq2_0}
\end{eqnarray}

The integrability conditions of equations (\ref{ce-eq1_0}-\ref{ce-eq2_0}), $\dif^2 p =0$ and $\dif^2 \xi =0$, can be written as:
\begin{eqnarray}
\dif \dot{p} \wedge u + (\theta \xi - 2 \dot{p})\, u \wedge a + \xi \, u \wedge m = 0 \label{ic_eq1_0}  \, ,\\
\dif \rho^{\!*} \! \wedge a + \xi \, \dif \theta \wedge u - \theta \rho^{\!*}\, u \wedge a + \rho^{\!*}\, u \wedge m =0  \, . \label{ic_eq2_0}
\end{eqnarray}
From equation (\ref{ic_eq2_0}) we can easily obtain:
\begin{equation} \label{ic_0}
\xi \, \dif \theta \wedge u \wedge a + \rho^{\!*}\, u \wedge m \wedge  a =0 \, .
\end{equation}
%

%
%
\subsection{Case $v \not=0$}
\label{subsec-w=0;vnot=0}

Now, condition (\ref{u-a-m-v}) applies, and $\{u,a,v,m\}$ is a basis. Then, (\ref{ic_0}) implies that  a scalar $\Upsilon$, concomitant of $u$, exists such that: 
\begin{equation} \label{Upsilon}
\rho^{\!*} = \Upsilon \, \xi \, , \qquad \Upsilon = \Upsilon[u] \equiv \frac{1}{v^2} *(\dif \theta \wedge u \wedge a \wedge v) \, .
\end{equation}
From here, we obtain $\dif \rho^* = \Upsilon \dif \xi + \xi \dif \Upsilon$. Then, making use of (\ref{ce-eq2_0}), equation (\ref{ic_eq2_0}) becomes:
\begin{equation} \label{ci_Upsilon}
\dot{p}\, \Upsilon \, u \wedge a = \xi \, [\Upsilon \, u \wedge m + \dif \theta \wedge u + \dif \Upsilon \wedge a] \, .
\end{equation}
%

%
\subsubsection{Subcase $\Upsilon \not=0$ {\rm (C$_{16}$).}}
\label{subsubsec-Ynot=0} 

\ \\[2mm]
Now, from (\ref{ci_Upsilon}) we can obtain $\Psi = \dot{p}/\xi$ as:
 \begin{equation} \label{chi-Ynot=0}
\hspace{-5mm} \Psi = \Psi_{16}[u] \equiv  \frac{1}{\Upsilon} i(u)i(a)[\dif \theta \wedge u + \dif(\Upsilon a)] = \frac{1}{a^2} (a,m) - \frac{1}{\Upsilon}(\dot{\Upsilon} + \theta^*) \, .
 \end{equation}
This expression determines $\Psi$ in terms of $u$ and, consequently, proposition \ref{prop-anot=0} applies:
\begin{proposition} \label{prop-Ynot=0}
A non-geodesic time-like unit vector $u$ with $w=0$, $v \not=0$ and $\Upsilon \not=0$, is the velocity of a perfect energy tensor if, and only if,  the pair $\{u, \Psi\}$, where $\Psi=\Psi_{16}[u]$ is given in {\em (\ref{chi-Ynot=0})}, fulfills equations {\em (\ref{eq-uchi})}.

Then, the energy density $\rho$ and the pressure $p$ are determined as proposition {\em \ref{prop-anot=0-inverse}} states.
\end{proposition}
%

%
\subsubsection{Subcase $\Upsilon=0$ {\rm (C$_{17}$).}}
\label{subsubsec-Y=0}

\ \\[2mm]
Under this condition, (\ref{Upsilon}) and (\ref{ci_Upsilon}) imply $\rho^* = 0$ and
 \begin{equation} \label{dtheta-u}
\dif \theta \wedge u = 0 \, .
 \end{equation}
Therefore, the constraints (\ref{duda-w=0}) mean that three independent functions $\{\tau, \alpha, \beta\}$ exist such that:
 \begin{equation} \label{tau,alpha,beta}
u = - e^{\alpha} \dif \tau \, , \qquad a = \dif \alpha + \beta\, \dif \tau   \, .
 \end{equation}
 Then, the conservation equations (\ref{ce-eq1_0}-\ref{ce-eq2_0}) are equivalent to:
 \begin{equation} \label{ce-tau,alpha}
\rho= \rho(\tau) \, , \qquad p_{,\alpha} + p + \rho = 0 , \qquad \rho'(\tau)= - \theta (\rho + p) e^{\alpha}   .
 \end{equation}
When $\theta \not=0$, the first and third equations above determine $\rho$ and $p$, respectively; and then, the second one is a consequence of (\ref{dtheta-u}). When $\theta =0$, we have $\rho= constant$, and the second equation enables us to obtain $p$. Consequently, we obtain:
\begin{proposition} \label{prop-Y=0}
A non-geodesic time-like unit vector $u$ with $w=0$, $v \not=0$ and $\Upsilon=0$, is the velocity of a perfect energy tensor if, and only if, $u$ fulfills $(u,v)=0$ and {\em (\ref{dtheta-u})}.

Then, two independent functions $\{\tau, \alpha\}$ exist such that $u = - e^{\alpha} \dif \tau$, and the energy density $\rho$ and the pressure $p$ are:
\begin{eqnarray}
\rho = \rho(\tau)  , \ \ \quad p = p(\tau,\alpha) = - \rho(\tau) - \frac{\rho'(\tau)}{\theta(\tau)}e^{-\alpha} , \qquad  
& \quad {\rm  if} \quad  \theta \not=0 \, , \label{rhop-thetanot=0} \\
\rho = \rho_0 , \ \ \qquad p = p(\tau,\alpha) = - \rho_0 - \varphi(\tau)e^{-\alpha} , \qquad  & \quad {\rm  if} \quad  \theta =0 \, , \label{rhop-theta=0}
\end{eqnarray}
where $\rho(\tau)$ and $\varphi(\tau)$ are arbitrary real functions.
\end{proposition}
%


\subsection{Case $v =0$ {\em (C$_{18}$)}}
\label{subsec-w=0;v=0}

Now, condition (\ref{u-a-m-v}) implies that $m = \zeta a$, that is $a\! \wedge\! m\!=\!0$, and then (\ref{ic_0}) leads to:
 \begin{equation} \label{dtheta-u-a}
\dif \theta \wedge u \wedge a= 0 \, .
 \end{equation}
Moreover, constraints (\ref{duda-w=0}) with $m = \zeta a$ imply that two independent functions $\{\tau, \beta\}$ and a function $\alpha = \alpha(\tau, \beta)$ exist such that:
\begin{equation} \label{tau,alpha,beta}
u = - e^{\alpha}  \dif \tau \, , \qquad a = \alpha_{,\beta} \, \dif \beta  \, , \qquad [{\rm and \ then} \ \ \zeta = - e^{-\alpha} (\ln \alpha_{,\beta})_{,\tau}]  \, .
 \end{equation}
Then, the conservation equations (\ref{ce-eq1_0}-\ref{ce-eq2_0}) are equivalent to:
 \begin{equation} \label{ce-tau,beta}
p_{,\beta} = - \alpha_{,\beta} (\rho+p)  , \qquad \rho_{,\tau}= -  \theta (\rho + p)e^{\alpha}   .
 \end{equation}
From these two equations we can obtain the following second-order partial differential equation for the function $p(\tau, \beta)$:
\begin{equation} \label{p-taubeta}
p_{,\tau \beta} = [(\ln \alpha_{,\beta})_{, \tau} - \theta\, e^{\alpha}]\, p_{, \beta} - \alpha_{,\beta}\, p_{,\tau}  .
\end{equation}
For every solution $p=p(\tau, \beta)$ to this equation, the first equation in (\ref{ce-tau,beta}) determines $\rho$, and then, the second one is an identity. Consequently, we obtain:
\begin{proposition} \label{prop-w=v=0}
A non-geodesic time-like unit vector $u$ with $w=0$ and $v=0$ is the velocity of a perfect energy tensor if, and only if, $u$ fulfills {\em (\ref{dtheta-u-a})}.

Then, two independent functions $\{\tau, \beta\}$ exist such that $u = - e^{\alpha} \dif \tau$ and $a = \alpha_{,\beta} \, \dif \beta$, and the energy density $\rho$ and the pressure $p$ are given by:
\begin{equation}
\rho = - p - \frac{p_{, \beta}}{\alpha_{, \beta}}  \, , \qquad p=p(\tau, \beta) \, ,  \label{rhode-p}
\end{equation}
where $p(\tau, \beta)$ is any solution of {\em (\ref{p-taubeta})}.
\end{proposition}
%

%
\begin{table}[t] \label{table-1}
\vspace*{-2.5mm}
\caption{This table shows the $u$-concomitants ${\cal D}_k[u]$ that we need to define the classes C$_n$ (second column) and those used to characterize each subset C$_n \cap {\bf U}_c$ (third column). We also indicate the equations where these concomitants are defined. The first column shows their order of differentiation.}
\noindent
\normalsize{
\begin{tabular}{lll} \label{table-1}
\\[-4mm]
\hline \\[-5.6mm] \hline
{\rm Order} & \ C$_n$-concomitants (eq.) &\phantom{\large $\frac{I^*}{I}\!$} \ S$_n$-concomitants (eq.)   \\
\hline \\[-5.6mm] \hline

1st  & \ $a$\,(\ref{acceleracio}),\ \ $\theta$\,(\ref{acceleracio}),\ \ $w$\,(\ref{rotations}),\ \ $b$\,(\ref{def-b}) &  \phantom{\large $\frac{I}{I}$}   \\

2nd & \ $v$\,(\ref{rotations}), \ \,$c$\,(\ref{def-c}), \ \,$\Omega$\,(\ref{cpi_alpha}), \ \,$\Upsilon$\,(\ref{Upsilon}) & \phantom{\large $\frac{I}{I}$} \ $\gamma$\,(\ref{def-c}), \ \,$z$\,(\ref{def-c}), \ \,$\Psi_2$\,(\ref{chi-C2})  \\

3rd & \ $\phi_1$ (\ref{phi_i}), \ \,$\phi_3$ (\ref{phi_i}), \ \,$\varphi_1$ (\ref{varphi_i})\ \,& \phantom{\large $ \ \frac{I}{I}$}\,$\Psi_3$ (\ref{chi-(cw)not=0}), \ \,$\Psi_{14}$\,(\ref{Omegues=0}), \ \,$\Psi_{16}$\,(\ref{chi-Ynot=0})  \\

4th & \ $h_1$\,(\ref{h_1}),\ $\kappa_1$\,(\ref{ic_kappa_i}),\ $\kappa_3$\,(\ref{ic_kappa_i}),\ $n_1$\,(\ref{n_1}) &   \phantom{\large $\frac{I}{I}$} \ $\phi_2$\ (\ref{phi_i}), \ \ $\Psi_{6}$\,(\ref{chi-6}), \ \ $\Psi_{11}$\,(\ref{Psi_alpha_tilde})\\


5th & \ $q_1$ (\ref{q_1}) & \phantom{\large $\frac{I}{I}$} \ $h_2$\,(\ref{h_1}), \ \ \,$\kappa_2$\,(\ref{ic_kappa_i}), \ \ \,$n_2$\,(\ref{n_2}), \\[-1mm]
&  &\phantom{\large $\frac{I}{I}$} \ $\Psi_{4}$ \ (\ref{g_tilde}), \ \ $\Psi_9$\,(\ref{kappa_1not=0}), \ \ $\Psi_{12}$\,(\ref{g_bar_3}) \\ 

6th &      &   \phantom{\large $\frac{I}{I}$} \ $q_2$\,(\ref{def-c}), \ \ $\Psi_7$\,(\ref{g_tilde_q_i})  \\
\hline \\[-5.6mm] \hline 
\end{tabular}
}
\vspace{-3mm}
\end{table}

\begin{table}[b] \label{table-2}
\vspace{-4mm}
\caption{The time-like unit vectors $u$ can be classified in eighteen classes C$_n$ ($n=1,...,18$) defined by relations imposed on the $u$-concomitants given in second column of table \ref{table-1}.}
\noindent
\normalsize{
\begin{tabular}{ll}  \label{table-2}
\\[-4mm]
\hline \\[-5.6mm] \hline
Classes & Definition relations \phantom{\large $\frac{I}{I}$} \\
\hline \\[-5.6mm] \hline

C$_1$ &   $a\! = \! 0$ \phantom{\large $\frac{I}{I}$}  \\

C$_2$ & $a\! \neq \! 0$, \quad   $(a,\!w)\!\not=\!0$  \phantom{\large $\frac{I}{I}$} \\

C$_3$ &  $a\! \neq \! 0$, \quad   $(a,\!w)\!=\!0$, \quad $(c,\!w)\!\not=\!0$ \phantom{\large $\frac{I}{I}$} \\

C$_4$ &  $a\! \neq \! 0$, \quad $w \! \not= \!0$, \quad  $(a,\!w)\!=\!(c,\!w)\!=\!0$, \quad $v\! \not=\! 0$, \quad 
$\phi_3\! \not= 0$, \quad  $\! h_1 \! \not= \!0$ \phantom{\large $\frac{I}{I}$} \\ 

C$_5$ & $a\! \neq \! 0$, \quad $w \! \not= \!0$, \quad  $(a,\!w)\!=\!(c,\!w)\!=\!0$, \quad $v\! \not=\! 0$, \quad 
$\phi_3\! \not= 0$, \quad  $\! h_1 \! = \!0$ \phantom{\large $\frac{I}{I}$} \\

C$_6$ & $a\! \neq \! 0$, \quad $w \! \not= \!0$, \quad  $(a,\!w)\!=\!(c,\!w)\!=\!0$, \quad $v\! \not=\! 0$, \quad 
$\phi_3\! = 0$, \quad  $\! \phi_1 \! \not= \!0$  \phantom{\large $\frac{I}{I}$}\\ 

C$_7$ & $a\! \neq \! 0$, \quad $w \! \not= \!0$, \quad  $(a,\!w)\!=\!(c,\!w)\!=\!0$, \quad $v\! \not=\! 0$, \quad 
$\phi_3\! = \! \phi_1 \!= \!0$, \quad   $\kappa_3 \! \not= \!0$, \quad   $q_1\! \not=\!0 \! \! \! \! \! \!$  \phantom{\large $\frac{I}{I}$}  \\

C$_8$ & $a\! \neq \! 0$, \quad $w \! \not= \!0$, \quad  $(a,\!w)\!=\!(c,\!w)\!=\!0$, \quad $v\! \not=\! 0$, \quad 
$\phi_3\! = \! \phi_1 \!= \!0$, \quad   $\kappa_3 \! \not= \!0$, \quad   $q_1\! =\!0 \! \! \! \! \! \!$  \phantom{\large$\frac{I}{I}$}  \\

C$_9$ &$a\! \neq \! 0$, \quad $w \! \not= \!0$, \quad  $(a,\!w)\!=\!(c,\!w)\!=\!0$, \quad $v\! \not=\! 0$, \quad 
$\phi_3\! = \! \phi_1 \!= \!0$, \quad   $\kappa_3 \! = \!0$, \quad   $\kappa_1\! \not=\!0 \! \! \! \! \! \!$ \phantom{\large $\frac{I}{I}$} \\

C$_{10}$ &  $a\! \neq \! 0$, \quad $w \! \not= \!0$, \quad  $(a,\!w)\!=\!(c,\!w)\!=\!0$, \quad $v\! \not=\! 0$, \quad 
$\phi_3\! = \! \phi_1 \!= \!0$, \quad   $\kappa_3 \! = \!0$, \quad   $\kappa_1\! =\!0 \! \! \! \! \! \!$ \phantom{\large $\frac{I}{I}$} \\

C$_{11}$ &  $a\! \neq \! 0$, \quad $w \! \not= \!0$, \quad  $(a,\!w)\!=\!(c,\!w)\!=\!0$, \quad $v\! =\! 0$, \quad 
$\tilde{c_b}\! +\! \tilde{\Omega} \! \not= \!0$  \phantom{\large $\frac{I}{I}$} \\ 

C$_{12}$ &  $a\! \neq \! 0$, \quad $w \! \not= \!0$, \quad  $(a,\!w)\!=\!(c,\!w)\!=\!0$, \quad $v\! =\! 0$, \quad 
$\tilde{c_b}\! =\! -\tilde{\Omega} \! \not= \!0$, \quad   $n_1\! \not=\!0 $ \phantom{\large $\frac{I}{I}$} \\

C$_{13}$ & $a\! \neq \! 0$, \quad $w \! \not= \!0$, \quad  $(a,\!w)\!=\!(c,\!w)\!=\!0$, \quad $v\! =\! 0$, \quad 
$\tilde{c_b}\! =\! -\tilde{\Omega} \! \not= \!0$, \quad   $n_1\! =\!0 $  \phantom{\large $\frac{I}{I}$}\\ 

C$_{14}$ & $a\! \neq \! 0$, \quad $w \! \not= \!0$, \quad  $(a,\!w)\!=\!(c,\!w)\!=\!0$, \quad $v\! =\! 0$, \quad 
$\tilde{c_b}\! =\! \tilde{\Omega} \! = \!0$, \quad \  \ $\varphi_1\not=0 $ \phantom{\large $\frac{I}{I}$}  \\

C$_{15}$ & $a\! \neq \! 0$, \quad $w \! \not= \!0$, \quad  $(a,\!w)\!=\!(c,\!w)\!=\!0$, \quad $v\! =\! 0$, \quad 
$\tilde{c_b}\! =\! \tilde{\Omega} \! = \!0$, \quad \  \ $\varphi_1=0 $ \phantom{\large $\frac{I}{I}$} \\

C$_{16}$ &  $a\! \neq \! 0$, \quad $w \! = \!0$, \quad   $v\! \not=\! 0$, \quad 
$\Upsilon\! \not=\!0$  \phantom{\large $\frac{I}{I}$} \\

C$_{17}$ &  $a\! \neq \! 0$, \quad $w \! = \!0$, \quad   $v\! \not=\! 0$, \quad 
$\Upsilon\! =\!0$ \phantom{\large $\frac{I}{I}$} \\ 

C$_{18}$ &  $a\! \neq \! 0$, \quad $w \! = \!0$, \quad   $v\! =\! 0$ \phantom{\large $\frac{I}{I}$} \\
\hline \\[-5.6mm] \hline
\end{tabular}
}
\vspace{-5mm}
\end{table}


%
\section{Velocities of a perfect energy tensor: summary theorems}
\label{sec-quadres}

In the previous sections we have shown that the characterization of the flow of a perfect fluid tensor requires considering a classification of the unit vectors $u$. The classes are defined by differential quantities associated with $u$. The second column in table \ref{table-1} shows these quantities, their differentiation order, and the equation in which they are defined. Similarly, the third column in table \ref{table-1} gives the same information about the quantities that are necessary to express the additional conditions characterizing the flows of a conservative energy tensor. For a vector $x$ and a scalar $\varphi$ we have considered the scalars defined in (\ref{x-base}) and (\ref{dphi-base}). With all this notation, we are led to introduce the following classification of the unit vector fields.
%

%
\begin{table}[t]
\vspace{-4mm}
\caption{The differential system S$_n$ in the second column gives the necessary and sufficient conditions for a
unit vector of class C$_n$ to belong to {\bf U}$_c$. The third column gives the expression H$_n$ of the pairs $\{\rho,p\}$ associated with a velocity in the set C$_n \cap {\bf U}_c$.}
\label{table-3}
\noindent
\normalsize{
\begin{tabular}{lll}
\\[-4mm]
\hline \\[-5.6mm] \hline
C$_n$ \ \,\ & S$_n$: nec.$\, $\&  suf.$\, $conditions&H$_n$: invers problem $\{\rho, p\}$ \phantom{\large $\frac{I^*}{I}$}  \\
\hline \\[-5.6mm] \hline 
\\[-5mm]
& \qquad \qquad \qquad &[$u = \dif \tau$] \qquad   $p \!=\! p(\tau)$, \quad  $\rho \ \ $ given in (\ref{rho-a=0-1}) \\[-2mm]
C$_1$ & $\emptyset$  & \\[-3mm]
 & \qquad \qquad \qquad     & [$u = \partial_\tau$] \qquad  \,$p= p_0$, \quad \,\, $\rho \ \ $ given in (\ref{rho-a=0-2})   \\[0mm]
\hline

C$_2$ & (\ref{eq-uchi}), \quad with $\ \Psi =\! \Psi_2[u]$\phantom{\large $\frac{I^2}{I}$}&  \\

C$_3$ & (\ref{eq-uchi}), \quad with $\ \Psi =\! \Psi_3[u]$\phantom{\large $\frac{I^2}{I}$}& \\

C$_4$ & (\ref{eq-uchi}), \quad with $\ \Psi =\! \Psi_4[u]$\phantom{\large $\frac{I^2}{I}$}&[$\ s = a + \Psi u, \quad
\Psi q = (\Psi - \theta ) s + i(u) \dif s\ $] \\ 

C$_6$ & (\ref{eq-uchi}), \quad with $\ \Psi =\! \Psi_6[u]$\phantom{\large $\frac{I^2}{I}$}&[$\ s = e^{\mu} \dif \lambda, \qquad \ q = \dif \mu \ $] \\ 

C$_7$ & (\ref{eq-uchi}), \quad with $\ \Psi =\! \Psi_7[u]$\phantom{\large $\frac{I^2}{I}$}&  \\

C$_9$ & (\ref{eq-uchi}), \quad with $\ \Psi =\! \Psi_9[u]$\phantom{\large $\frac{I^2}{I}$}&
$p = - \lambda, \ \  \rho = -p + e^{-\mu}$  \qquad\,\quad {\rm  if} \ \,$\Psi \not=0$  
 \\

C$_{11}$ & (\ref{eq-uchi}), \quad with $\ \Psi =\! \Psi_{11}[u]$\phantom{\large $\frac{I^2}{I}$}&
$p\! =\! p(\lambda),  \ \, \rho\! =\! -p(\lambda)\! -\! p'(\lambda) e^{-\mu}$\ \,    {\rm  if} \ \,$ \Psi =0$    \\ 

C$_{12}$  & (\ref{eq-uchi}), \quad with $\ \Psi =\! \Psi_{12}[u]$\phantom{\large $\frac{I^2}{I}$}& \\

C$_{14}$ & (\ref{eq-uchi}), \quad with $\ \Psi =\! \Psi_{14}[u]$\phantom{\large $\frac{I}{I}$}&  \\

C$_{16}$ & (\ref{eq-uchi}), \quad with $\ \Psi =\! \Psi_{16}[u]$\phantom{\large $\frac{I^2}{I}$}& \\
\hline

C$_5$ & $(u,v)=\hat{\Omega} \!-\! \gamma z_w\! =\!h_2\! =\!0$ \phantom{\large $\frac{I^2}{I}$}& \\

C$_8$ & $(u,v)=\hat{\Omega} \!-\! \gamma z_w\! =\!\phi_2\! =\!q_2 = 0$  \phantom{\large$\frac{I^2}{I}$}&see Appendix A  \\

C$_{13}$ &$\hat{\Omega}= \tilde{\tilde{\Omega}} + \tilde{\Omega}\, c_b = n_2 = 0$  \phantom{\large $\frac{I^2}{I}$}& \\  
\hline

C$_{10}$ &  $(u,v)=\hat{\Omega} \!-\! \gamma z_w\! =\!\phi_2\! =\!\kappa_2\! =\!0$ \phantom{\large $\frac{I^2}{I}$}&see Appendix B \\

C$_{15}$ & $\hat{\Omega} = \dot{\Omega} + 2 \Omega c_b = 0$ \phantom{\large $\frac{I^2}{I}$}&$\{\rho, p\}$ given in (\ref{rhode-p-C}) \\

C$_{17}$ &  $(u,v)=\dif \theta \wedge u  =  0$\phantom{\large $\frac{I^2}{I}$}&$\{\rho, p\}$ given in (\ref{rhop-thetanot=0}-\ref{rhop-theta=0}) \\

C$_{18}$ &   $\dif \theta \wedge u \wedge a  =  0$\phantom{\large $\frac{I^2}{I}$}&$\{\rho, p\}$ given in (\ref{rhode-p}) \\
\hline \\[-5.6mm] \hline
\end{tabular}
}
\vspace{-5mm}
\end{table}


%
\begin{definition}
{\em ({\bf classification of the flows})}
A time-like unit vector $u$ is said to be of class {\em C$_n$ ($n=1,...,18$)} if it satisfies the relations given in table {\em \ref{table-2}}, where the involved $u$-concomitants are tabulated in the second column of table {\em \ref{table-1}}.
\end{definition}

Then, the results in the above sections can be summarized by the following two theorems:
\begin{theorem}
{\em ({\bf characterization of the conserved flows})} A time-like unit vector $u$ of class {\em C$_n$ ($n=1,...,18$)}
is the velocity of a perfect energy tensor if, and only if, it satisfies the differential system {\em S$_n$} given in the second column of table {\em \ref{table-3}}, where the involved $u$-concomitants are tabulated in the third column of table {\em \ref{table-1}}. 
\end{theorem}
\begin{theorem}
The pairs $\{\rho,p\}$ of hydrodynamic quantities associated with a velocity of class {\em C$_n$ ($i=1,...,18$)} that fulfills the condition {\em S$_n$} are determined by the expressions {\em H$_n$} given in the third column of table {\em \ref{table-3}}.
\end{theorem}


\noindent

\begin{figure}
\hspace*{10mm} 
\vspace*{-28mm}
\setlength{\unitlength}{0.9cm} {\small \noindent
\begin{picture}(0,18)
\thicklines

\put(4.7,17){\line(-3,-1){1.2}}
 \put(2.3,17){\line(3,-1){1.2}}
\put(4.7,17){\line(0,1){0.6}} \put(2.3,17.6){\line(1,0){2.4}}
\put(2.3,17.6){\line(0,-1){0.6}}

 \put(2.7,17.05  ){$ u, \   {\cal D}_k [u]  $}

\put(3.5,16.60){\vector(0,-1){0.7 }}

\put(3.5,15.9){\line(-2,-1){1 }} \put(3.5,15.9 ){\line(2,-1){1 }}
\put(3.5,14.9){\line(2,1){1 }} \put(3.5,14.9){\line(-2,1){1 }}
\put(3,15.25){$a=0 $}

 \put(2.5,15.4){\vector(-1,0){2.9}}
\put(-1.3,15.25){\framebox{\,C$_1$\,}}
 \put(4.5,15.4){\vector(1,0){1.25}}

\put(7,16 ){\line(-2,-1){1.25}} \put(7,16  ){\line(2,-1){1.25}}
\put(7,14.75){\line(2,1){1.25}} \put(7,14.75){\line(-2,1){1.25}}
\put(6.2,15.25){$(a ,\!w)\!=\!0 $}

\put(7,14.25 ){\line(-2,-1){1 }} \put(7,14.25){\line(2,-1){1 }}
\put(7,13.25){\line(2,1){1 }} \put(7,13.25){\line(-2,1){1 }}
\put(6.5,13.6){$w\neq0 $}

\put(1.75,14.25 ){\line(-2,-1){1 }} \put(1.75,14.25){\line(2,-1){1
}} \put(1.75,13.25){\line(2,1){1 }} \put(1.75,13.25){\line(-2,1){1
}} \put(1.25,13.6){$v\neq 0 $}

 \put(0.75,13.75){\vector(-1,0){1.13}}
 \put(0.75,12.25){\vector(-1,0){1.12}}
\put(-1.3,13.65){\framebox{C$_{18}$}}
 \put(-1.3,12.15){\framebox{C$_{17}$}}

 \put(6,13.75){\vector(-1,0){3.3}}

\put(1.75, 11.75){\vector(0,-1){0.5}}
 \put(1.75,11.25){\vector(-1,0){2.12}}



\put(1.75, 13.25){\vector(0,-1){0.5}}
 \put(-1.3,11.15){\framebox{C$_{16}$}}

\put(1.75,12.75 ){\line(-2,-1){1 }} \put(1.75,12.75){\line(2,-1){1
}} \put(1.75,11.75){\line(2,1){1 }} \put(1.75,11.75){\line(-2,1){1
}} \put(1.25,12.1){$\Upsilon\!\neq0 $}

 \put(7,12.9 ){\line(-2,-1){1.25}}
 \put(7,12.9){\line(2,-1){1.25}} \put(7,11.65){\line(2,1){1.25}}
\put(7,11.65){\line(-2,1){1.25}} \put(6.2,12.15){$(c ,\!w)\!=\!0 $}

\put(7,11.25 ){\line(-2,-1){1 }} \put(7,11.25){\line(2,-1){1 }}
\put(7,10.25){\line(2,1){1 }} \put(7,10.25){\line(-2,1){1 }}
\put(6.5,10.6){$v=0 $}


\put(9.5,10.25 ){\line(-2,-1){1.25}}
\put(9.5,10.25){\line(2,-1){1.25}} \put(9.5,9){\line(2,1){1.25}}
\put(9.5,9){\line(-2,1){1.25}}
\put(8.67,9.475){$\tilde{c_b}\!+\!\tilde{\Omega}\! =\!0 $}

\put(9.5,8.5 ){\line(-2,-1){1.25}} \put(9.5,8.5){\line(2,-1){1.25}}
\put(9.5,7.25){\line(2,1){1.25}} \put(9.5,7.25){\line(-2,1){1.25}}
\put(8.9,7.7){$\tilde{\Omega}=0 $}

\put(9.5,6.2 ){\line(-2,-1){1.25}} \put(9.5,6.2){\line(2,-1){1.25}}
\put(9.5,4.95){\line(2,1){1.25}} \put(9.5,4.95){\line(-2,1){1.25}}
\put(8.85,5.47){$\varphi_1=0 $}

 \put(10.7, 5.6){\vector(1,0){4.06}}

\put(14.75,5.49){\framebox{C$_{14}$}}

\put(14.75,4.39){\framebox{C$_{15}$}}

 \put(12.75,8.37 ){\line(-2,-1){1 }}
\put(12.75,8.37){\line(2,-1){1 }} \put(12.75,7.37){\line(2,1){1 }}
\put(12.75,7.37){\line(-2,1){1 }} \put(12.25,7.75){$n_1\!=\!0 $}

\put(14.75,7.69){\framebox{C$_{12}$}}

\put(14.75,6.6){\framebox{C$_{13}$}}

 \put(9.5, 9){\vector(0,-1){0.5}}

\put(10.75, 9.63){\vector(1,0){4}}

 \put(14.75,9.56){\framebox{C$_{11}$}}

\put(10.75, 7.89){\vector(1,0){1}}

\put(13.75, 7.89){\vector(1,0){1}} \put(12.75, 6.7){\vector(1,0){2}}

\put(12.75 , 7.4){\vector(0,-1){0.7}}

\put(9.5 , 7.27){\vector(0,-1){1.08}}

\put(9.5 , 4.96){\vector(0,-1){0.5}}

\put(9.5 , 4.46){\vector(1,0){5.26}}

\put(14.75,15.25){\framebox{\,C$_2$\,}}
 \put(8.2,15.4){\vector(1,0){6.55}}
\put(14.75,12.2){\framebox{\,C$_3$\,}}
\put(8.2,12.27){\vector(1,0){6.55}}

\put(3.75 , 9.76){\vector(-1,0){1}}

\put(0.75 , 9.76){\vector(-1,0){1.17}}

\put(-1.3,9.7){\framebox{\,C$_4\, $}}

\put(1.75,9.25){\vector(0,-1){0.5}}
\put(1.75,8.76){\vector(-1,0){2.16}}

\put(-1.3,8.6){\framebox{\,C$_5$\,}}


\put(4.75,10.25 ){\line(-2,-1){1 }} \put(4.75,10.25){\line(2,-1){1
}} \put(4.75,9.25){\line(2,1){1 }} \put(4.75,9.25){\line(-2,1){1 }}
\put(4.2,9.63){$\phi_3\!=\!0 $}

\put(1.75,10.25 ){\line(-2,-1){1 }} \put(1.75,10.25){\line(2,-1){1
}} \put(1.75,9.25){\line(2,1){1 }} \put(1.75,9.25){\line(-2,1){1 }}
\put(1.25,9.63){$h_1\!=\!0 $}

\put(4.75, 10.75){\vector(0,-1){0.5}}

\put(4.75, 10.75){\line(1,0){1.3}}

\put(8, 10.75){\line(1,0){1.5}}

\put(9.5, 10.75){\vector(0,-1){0.5}}

\put(3.8,6.25){\vector(-1,0){1.1}}

\put(3.8,7.85){\vector(-1,0){4.2}}

 \put(4.75,8.35 ){\line(-2,-1){1 }} \put(4.75,8.35){\line(2,-1){1}}
 \put(4.75,7.35){\line(2,1){1 }} \put(4.75,7.35){\line(-2,1){1 }}
\put(4.2,7.73){$\phi_1\!=\!0 $}

\put(-1.3,7.7){\framebox{\,C$_6$\,}}


\put(4.75,6.75 ){\line(-2,-1){1 }} \put(4.75,6.75){\line(2,-1){1}}
 \put(4.75,5.75){\line(2,1){1 }} \put(4.75,5.75){\line(-2,1){1 }}
\put(4.2,6.13){$\kappa_3\!=\!0 $}

\put(1.75,6.75 ){\line(-2,-1){1 }} \put(1.75,6.75){\line(2,-1){1 }}
\put(1.75,5.75){\line(2,1){1 }} \put(1.75,5.75){\line(-2,1){1 }}
\put(1.25,6.13){$q_1\!=\!0 $}

\put(1.75,5.75){\vector(0,-1){0.5}}

\put(1.75,5.25){\vector(-1,0){2.16}}

\put(-1.3,6.2){\framebox{\,C$_7$\,}}

 \put(-1.3,5.1){\framebox{\,C$_8$\,}}

\put(0.75,6.25){\vector(-1,0){1.16}}

 \put(-1.3,4){\framebox{\,C$_9$\,}}

\put(-1.3,2.9){\framebox{C$_{10}$}}

\put(3.75,4.15){\vector(-1,0){4.16}}
\put(4.75,3){\vector(-1,0){5.13}}

\put(4.75,3.65){\vector(0,-1){0.65}}

\put(4.75,4.65 ){\line(-2,-1){1 }} \put(4.75,4.65){\line(2,-1){1}}
 \put(4.75,3.65){\line(2,1){1 }} \put(4.75,3.65){\line(-2,1){1 }}
\put(4.2,4.03){$\kappa_1\!=\!0 $}

\put(4.75,5.75){\vector(0,-1){1.1}}

\put(4.75,7.35){\vector(0,-1){0.6}}

\put(4.75,9.3){\vector(0,-1){0.94}}

\put(7,11.7){\vector(0,-1){0.45}}

\put(7,13.25){\vector(0,-1){0.35}}

\put(7,14.75){\vector(0,-1){0.5}}


\put(1.4,15.5){yes}

\put(4.8,15.5){no} \put(9,15.5){no}

\put(9,12.4){no}

\put(7.2,14.45){yes}

\put(4.8,13.9){no}

\put(0.1,13.9){no}

\put(0.1,12.4){no}

\put(1.9,11.4){yes}

\put(7.2,12.95){yes}

\put(7.2,11.35){yes}

\put(8.4,10.9){yes}

\put(5.3,10.9){no}

\put(0.1,9.9){no}

\put(3.1,9.9){no} \put(11.3,9.75){no}

\put(1.9,8.95){yes}

\put(4.9,8.9){yes}

\put(9.68,8.7){yes}

\put(3.1,8){no}

\put(10.8,8){no}

\put(13.8,8){no}

\put(12.9,7){yes}\put(9.65,6.8){yes} \put(4.95,7){yes}

\put(3.2,6.35){no} \put(0.1,6.35){no}

\put(1.93,5.4){yes} \put(1.93,12.95){yes}

\put(4.93,5.4){yes} \put(4.93,3.23){yes} \put(3.2,4.23){no}

\put(10.9,5.8){no}

\put(9.75,4.7){yes}

\end{picture} }
\vspace{2mm}
\caption{This flow diagram distinguishes the different classes of fluxes of perfect energy tensors.}
\label{figure-1}
\end{figure}
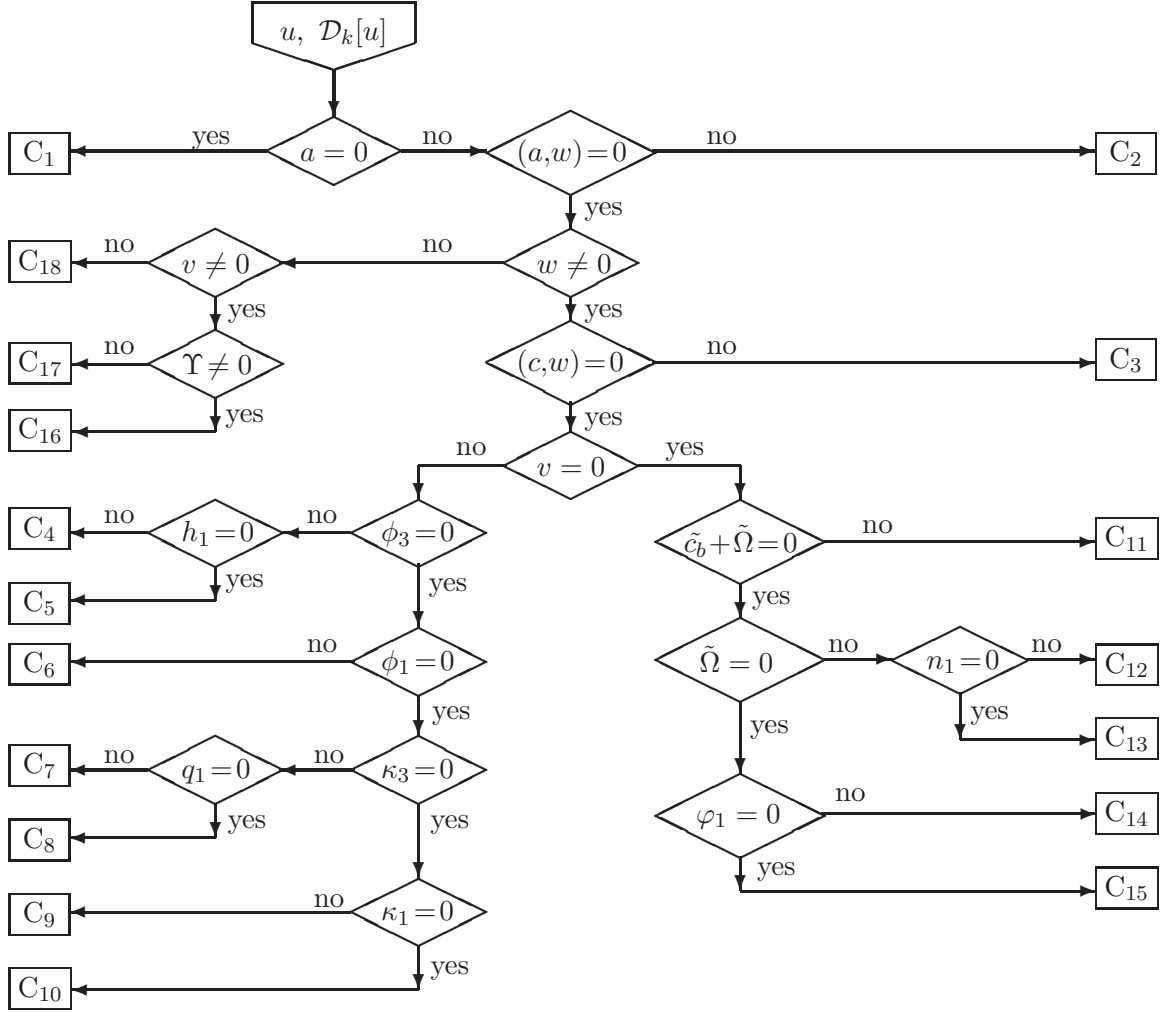
\vspace{0mm}


It is worth remarking that the results of the theorems above offer an IDEAL characterization of the velocities of a perfect energy tensor. This means that an algorithm can be built that enables us to distinguish every class C$_n$ and to test the labeling conditions S$_n$. We present this algorithm as a flow diagram (see figure \label{figure}). The input data are $u$ itself and the differential $u$-concomitants ${\cal D}_k[u]$ presented in table \ref{table-1}. If a velocity belongs to class C$_n$, it must fulfill the necessary and sufficient conditions S$_n$ in order to be the velocity of a perfect energy tensor.


\section{Some examples}
\label{sec-examples}

In this section, we want to highlight the interest of the theoretical results presented in this paper by analyzing some examples with our approach of both test solutions and self-gravitating systems. We only stress some first outcomes that  we will tackle in further work.


\subsection{Perfect fluids with a stationary flow}
\label{subsec-Killing}

Let $u$ be a unit velocity such that $\xi = |\xi| u$ is a Killing vector. Then, $\theta = 0$, $\sigma =0$, and  $\dif a = 0$, where $\sigma$ is the shear.
Moreover $a = \dif \alpha$, where $\alpha = \ln |\xi|$. Then, the study of the compatibility of these kinematic restrictions with our classification shows that only classes C$_1$, C$_2$, C$_{15}$ and C$_{18}$ are compatible. Moreover, our study of these classes enables us to state:
\begin{proposition} \label{propo-killing} 
Let $\xi$ be a time-like Killing vector, then $u = \xi/|\xi|$ is always the unit velocity of a perfect energy tensor.

Case $a=0$ {\em [C$_1$]}. (i) if $w\not=0$, then the pressure is an arbitrary constant, $p=p_0$; (ii) if $w=0$, then $u=- \dif \tau$ and $p=p(\tau)$. In both cases the energy density is an arbitrary $u$-invariant function, $\rho = \rho(\varphi_i)$, $\dot{\varphi_i}=0$, $i=1,2,3$.

Case $(w,a)\not=0$ {\em [C$_2$]}. Then, the pressure is an arbitrary function of the norm of $\xi$, $p=p(\alpha)$, $\alpha = \ln |\xi|$, and the energy density is given by $\rho = \rho(\alpha) \equiv -p(\alpha) - p'(\alpha)$.

Case $a \not=0, \, w \not=0,\, (w,a)=0$ {\em [C$_{15}$]}. Then, then $u=- \dif \tau + \mu \dif \alpha$, and the pressure and the energy density are given, respectively, by $p= c_1\tau e^{-2\alpha} + p_1(\alpha)$ and $\rho =c_1(\tau+\mu) e^{-2\alpha} -p_1(\alpha) - p_1'(\alpha)$.

Case $a \not=0, \, w =0$ {\em [C$_{18}$]}. Then, $u=-  e^{-\alpha}\dif \tau$, and the pressure and the energy density are, respectively, by $p=p_0(\tau) e^{-\alpha} + p_1(\alpha)$ and $\rho = -p_1(\alpha) - p_1'(\alpha)$.
\end{proposition}
If we have a stationary flow, then $u$ fulfills equations (\ref{eq-chi=0}) and the conservation equations admit isobaric ($\dot{p}=0$) solutions. This agrees with the above statement. Moreover, in this case we have a barotropic evolution, $d \rho \wedge d p=0$. However, in cases C$_{1}$, C$_{15}$ and C$_{18}$ there are also nonisobaric solutions. But these solutions cannot be interpreted as perfect fluids evolving in local thermal equilibrium (see next subsection). 

Note that the observer at rest with respect to a non-geodesic static gravitational field defines a flow of type C$_{18}$, and the associated energy tensor is stationary when $p_0(\tau) = constant$. In this case, each choice of the function $p_1(\alpha)$ defines a specific barotropic relation $p=p(\rho)$. A test fluid at rest in the Schwarzschild spacetime or a self gravitating sphere in equilibrium belongs to this class. 

However, the perfect fluids at rest in the Kerr solution or in any stationary axisymmetric gravitational field are, generically, of class C$_2$. The determination of the subset of stationary axisymmetric solutions with a flow of class C$_{15}$ is an open problem that will be considered elsewhere. Further work is required to study a similar question for the flows defined by a conformal Killing vector field.


\subsection{Radial flow in a spherically symmetric spacetime}
\label{subsec-caiguda radial}

Test fluids with a radial flow can model different astrophysical and cosmological scenarios. For example, the radial accretion of a fluid onto a spherically symmetric central object has been used as a simple but non-trivial accretion model to study the effects of backreaction (see \cite{schn-baug-shap, mach-malec} and references therein). 

A spherically symmetric metric can be written in the general form $\dif s^2 = -e^{2\nu (t,r)} \dif t^2 + e^{2\lambda(t,r)} \dif r^2+ R^2(t,r) (\dif \vartheta^2+  \sin ^2 \vartheta \dif \varphi^2)$, while a fluid radially moving has a unit velocity of the form:
\begin{equation}
u = - e^{\nu}\cosh \psi \, \dif t +e^{\lambda}  \sinh \psi \, \dif r \, , \qquad \psi = \psi(t,r) \, .
\end{equation}
Then, the fluid acceleration takes the expression: 
\begin{equation}
\hspace{-20mm}a = A(t,r)[-e^{-\lambda}\!  \sinh \psi \, \dif t \!+\! e^{-\nu}\!\cosh \psi \, \dif r]  , \quad  A(t,r) \!\equiv \! (e^{\lambda}\!  \sinh \psi)_{,t}\! +\! (e^{\nu}\! \cosh \psi)_{,r}   .
\end{equation}

If the flow is geodesic (class C$_1$), the function $\psi(t,r)$ fulfills the first-order partial differential equation $A(t,r)=0$, which always admits solution. 
Otherwise, if $a \not=0$, we have $a \wedge \dif a = u \wedge \dif u = 0$, and it necessarily belongs to class C$_{18}$. In both cases, such a $u$ defines the flow of a solution to the conservation equation (\ref{conservation-1}).

The solution $\{\rho, p\}$ to the inverse problem depends on the specific metric functions $\nu(t,r)$ and $\lambda(t,r)$. For example, in the Minkowski spacetime, $\nu=\lambda=0$, and for a geodesic flow, equation $A(t,r)=0$ has the general solution $t \!-\! r \coth \psi \!=\! f(\psi)$, where $f(\psi)$ is an arbitrary real function. The particular solution $f(\psi) = t_0$ leads to the Milne's flow, and then $u= - \dif \tau$, $\tau = [1- z^2]^{-1/2}$, $z \equiv r/(t-t_0)$. In this case, $\dif \theta \wedge u =0$, and according to a result in \cite{CFS-CIG}, $u$ is the flow of a classical ideal gas. It is also the flow of the perfect energy tensors defined by the solution (\ref{rho-a=0-1}) of the inverse problem.


\subsection{On the Von Zeipel theorem and its extensions}
\label{subsec-vonzeipel}

Let us consider a stationary axisymmetric spacetime with a metric line element of the form $\dif s^2 = g_{tt} \dif t^2 + 2 g_{t \varphi} \dif t \dif \varphi + g_{rr} \dif r^2+ g_{\vartheta \vartheta} \dif \vartheta^2+ g_{\varphi \varphi} \dif \varphi^2$, $g_{\alpha \beta} = g_{\alpha \beta}(r, \vartheta)$, where $(t, r, \vartheta,\varphi)$ are spherical-like coordinates and $X = \partial_t$ and $Y = \partial_{\varphi}$ are Killing vectors. A stationary axisymmetric perfect fluid with purely circular flow has a unit velocity of the form \cite{ReZa}: 
\be \label{von-zeipel-a}
\hspace{-10mm} u = (u^{\alpha}) = u^t (\partial_t + \Omega  \partial_{\varphi}) \,  , \quad (u_{\alpha}) = u_t (\dif t - \ell \dif \varphi) \, ,  \quad u^t u_t = [\Omega \ell -1]^{-1} ,
\ee
where $\Omega$ is the {\em coordinate angular velocity} and $\ell$ the {\em specific angular momentum}. 

The so-called relativistic Von Zeipel theorem \cite{ReZa, Abramowicz} states that the perfect fluid is barotropic if, and only if, the hypersurfaces of constant $\Omega$ coincide with the hypersurfaces of constant $\ell$. Note that the first is a hydrodynamic property, $\dif \rho \wedge \dif p =0$, while the second one is a purely kinematic condition, $\dif \Omega \wedge \dif \ell =0$. 

This fact suggests that our kinematic characterization could bring new insights into the Von Zeipel theorem.  A straightforward calculation leads to:
\be \label{von-zeipel-b}
u \wedge \dif u = u_t^2 \, \dif t \wedge \dif \varphi \wedge \dif \ell \, , \qquad  \dif a = \ell [\Omega \ell -1]^{-2} \dif \ell \wedge \dif \Omega \, .
\ee
Thus, the kinematic constraint $\dif \Omega \wedge \dif \ell =0$ is equivalent to the intrinsic kinematic condition $\dif a =0$. Note that a stationary fluid implies $\dot{p}=0$. Then, the flow is geodesic (class C$_1$) if, and only if, $p = constant$. Moreover, the first equation in (\ref{von-zeipel-b}) implies that the flow is irrotational ($w=0$) if, and only if, the specific angular momentum is constant ($\dif \ell =0$), and then $u$ belongs to class C$_{18}$. Otherwise, when $w \not= 0$, the flow belongs to class C$_{15}$.  In these two cases, the inverse problem (see propositions \ref{prop-w=v=0} and \ref{prop-kappa_i=0}, respectively) leads to a pressure and an energy density given by:
\be \label{von-zeipel-c}
p= p(\alpha) \, , \qquad  \rho (\alpha) = - p(\alpha) - p'(\alpha)  \, ,
\ee
where $p(\alpha)$ is an arbitrary real function of the acceleration potential $\alpha$, $\dif \alpha = a$. Then, we have $\dif \rho \wedge \dif p = 0$, and we recover the Von Zeipel theorem. 

The comprehensive study of the unit velocities of the form (\ref{von-zeipel-a}) taking into account the results of this work shows that all these vector fields define the flow of a conservative perfect fluid, and that, in addition to the classes quoted above, only classes C$_2$ and C$_3$ are possible. The analysis of the inverse problem for these two classes and the study of the compatible nonbarotropic equations of state could lead to potential extensions of the Von Zeipel theorem. 


\subsection{The flow of the Stephani universes}

In the three previous examples, we have considered test fluids in a given gravitational field. The analysis of the kinematics of a self-gravitating perfect fluid taking into account our approach can also be of interest. The divergence-free condition holds as a consequence of the field equations but, depending on the class C$_n$, the inverse problem can provide test perfect fluids that are comoving with the self-gravitating system. As an example, we analyze the Stephani universes.

The conformally flat perfect fluid solutions to Einstein equations with non-zero expansion were obtained by Stephani \cite{Stephani}. They were rediscovered \cite{Barnes} as the conformally flat class of expanding, irrotational and shear-free perfect fluid space\-times. The metric line element takes the expression $\dif s^2 = -\alpha^2 dt^2 + \Omega^2 \delta_{ij} dx^i dx^j$, where $\alpha \equiv R\, \partial_R \ln \Omega$ and $\Omega \equiv R(t)[1+ 2 \vec{b}(t) \cdot \vec{{\rm r}} + \frac{1}{4} K(t) r^2]^{-1}$, and the unit velocity of the fluid is $u = \alpha^{-1} \partial_t$. Moreover, the energy density, pressure and expansion are:
\begin{equation}
\hspace{-10mm} \rho = \frac{3}{R^2}(\dot{R}^2 + K - 4b^2), \qquad p= - \rho - {R \over 3} {\partial_R\rho \over \alpha}, \qquad \theta(t) = {3 \dot{R} \over R} \not= 0.  \label{eq:dp}
\end{equation}
The geodesic case (class C$_1$) leads to the LFRW limit. The strict Stephani universes ($a \not=0$) have an irrotational flow ($w=0$) with homogeneous expansion ($\dif \theta \wedge u = 0$). Consequently, only classes C$_{17}$ and C$_{18}$ are compatible.

In class C$_{17}$ ($v \not=0$) the inverse problem (see proposition \ref{prop-Y=0}) leads to an arbitrary homogeneous energy density, $\bar{\rho} = \bar{\rho}(t)$ (which is not given by the first expression in (\ref{eq:dp})), and a pressure $\bar{p}$ given by the second expression in (\ref{eq:dp}) replacing $\rho$ by $\bar{\rho}$.

Besides, it can be shown that condition $v =0$ (class C$_{18}$) leads to the Stephani universes admitting a G$_3$, that is, those representing the evolution of a fluid in local thermal equilibrium \cite{Bona-Coll}. The study of the inverse problem for this class, and for the subclass that can be interpreted as generic ideal gas \cite{C-F}, will be considered elsewhere.  


\section{Discussion}
\label{sec-discussion}

The precise description of the kinematics of the relativistic continuous media was performed years ago by Ehlers \cite{Ehlers}. In this general approach, the flow of the media is described by a time-like unit vector field, and the kinematic properties depend on its first order derivatives, collected in the acceleration and rotation vectors $a$ and $w$, the expansion scalar $\theta$ and the shear tensor $\sigma$. Then, it is natural to classify the fluids according to the nullity or not of these kinematic quantities. The subsequent classes do not depend on the hydrodynamic equations nor on the existence of viscidity or heat flux. Consequently, any time-like unit vector belongs to one of the classes. 

Our approach here is quite different. We are interested in the fluid flow of the divergence-free perfect energy tensors and we have shown that this fact imposes constraints on the time-like unit vectors. It is just the study of these restrictions that leads to the eighteen classes considered in this paper. It is worth mentioning that conditions C$_n$ defining the classes and equations S$_n$ characterizing the fluxes involve the kinematic quantities $a$, $w$, $\theta$ and their derivatives, but the shear $\sigma$ plays no role in our study. We can find a similar situation in analyzing the barotropic flows \cite{CF-barotrop} and in characterizing the kinematics of a classical ideal gas \cite{FS-KCIG}. 

The full development of possible applications of our results in the search for solutions of the hydrodynamic system that model a specific physical system and the exhaustive analysis of other applications, fall outside the scope of this work. In section \ref{sec-examples}, we have outlined some issues where our approach could be useful and that require further studies with more detail: the analysis of the fluids at rest in the Kerr (or other stationary) gravitational field, the study of the flow defined by a conformal Killing field, the better understanding of the radial infall in Schwarzschild or in other spherically symmetric spacetimes, the extension of the Von Zeipel theorem, and the classification of the flow of significant perfect fluid solutions of Einstein equations together with the obtention of the test fluids comoving with the perfect fluid source. Similarly, our kinematic approach can be useful to analyze other physical systems that may be of interest in astrophysics and cosmology. Finally, a subject in which we are working at the moment is that of thermodynamic flows.

\subsection{Thermodynamic flows}
\label{subsec-termo}

If we are interested in modeling a perfect fluid in local thermal equilibrium, in addition to the {\em hydrodynamic quantities} $\{u, \rho, p\}$ subjected to the conservation equations (\ref{conservation-1}) we must consider a set of {\em thermodynamic quantities} $\{n, s ,\Theta\}$ (matter density $n$, specific entropy $s$ and temperature $\Theta$) constrained by the usual thermodynamic laws \cite{Eckart}, namely, the conservation of matter, and the local thermal equilibrium relation. 

In \cite{CFS-LTE} we have shown that a thermodynamic non-barotropic evolution of a perfect fluid can be expressed in terms of the sole hydrodynamic quantities $\{u, \rho, p\}$: if the evolution is isoenergetic ($\dot{\rho}=0$), the local thermal equilibrium requires an isobaric evolution ($\dot{p}=0$); and if $\dot{\rho}\not=0$, a thermodynamic evolution can be characterized by the {\em hydrodynamic sonic condition}, $\dif \chi \wedge \dif \rho \wedge \dif p = 0$, where $\chi = \dot{p}/\dot{\rho}$ is the {\em indicatrix function} \cite{CFS-LTE}. In this case, $\chi$ is a function of state, which represents the square of the speed of sound in the fluid, $\chi = \chi(\rho, p) = c_s^2$ (see \cite{CFS-regular} for more details). 

We can ask ourselves if a unit time-like vector $u$ defines a {\em thermodynamic flow}, that is, if it is the velocity of a conservative prefect fluid fulfilling the hydrodynamic sonic condition. Clearly, such a $u$ belongs to {\bf U}$_c$, and consequently, the results in this paper should provide an essential tool for solving this open problem. First, we must study which classes C$_n$ are compatible with the hydrodynamic sonic condition, and second, we must determine, for each class, the subset of velocities defining a thermodynamic flow. These issues will be discussed elsewhere.

It is worth remarking that such a study was accomplished for the barotropic fluids years ago \cite{CF-barotrop}, and more recently for the classical ideal gases \cite{CFS-CIG}. Analogous research has yet to be done for the generic ideal gases, that is, when the indicatrix function is of the form $\chi= \chi(p/\rho)$ \cite{CFS-LTE}.


\appendix

\section{\ }
\label{ap-A}

Solving the inverse problem for class C$_5$ (respectively, C$_8$ and C$_{13}$) requires studying the exterior system (\ref{dpdot_dxi}) (respectively, (\ref{dpdot_dxi_n_i}) and (\ref{dpdot_dxi_gamma=0})) when the integrability conditions hold identically. These conditions give:
\begin{equation} \label{dosr}
\hspace{-15mm}\dif g_1 = g_2 \wedge g_3, \ \ \dif g_2 = g_2 \wedge (g_4 -g_1) , \ \ \dif g_3 =(g_4 -g_1)
\wedge g_3 , \ \ \dif g_4 =  g_3 \wedge g_2 . 
\end{equation}
Let us consider  $\xi = \zeta \, \dot{p} $. Then, the system (\ref{dpdot_dxi}) is equivalent to
\begin{equation} \label{A2}
\dif \dot{p} =\dot{p}  \, ( g_1 +\zeta \, g_2 ) \, , \qquad 
\dif \zeta = g_3 + \zeta (g_4 -g_1) - \zeta^2 g_2 \, . \label{ap1e}
\end{equation}
Moreover, the exterior system (\ref{dosr}) implies that two functions $\{\alpha, \beta\}$ exist such that
\be \label{A3} 
g_2=e^{\alpha} \dif \beta \, , \qquad 2 \zeta_0 \,g_2+ g_1 -g_4 = \dif \alpha \, ,
\ee
where $\zeta_0$ is a particular solution of the second equation in (\ref{ap1e}). Then, we can determine the general solution $\zeta$ of this Riccati-like equation, and we have:
\be \label{A4}
\zeta = \zeta_0 + \frac{c_0 e^{-\alpha}}{\beta + c_1}   \, ,  \qquad g_1+ \zeta \,g_2  = \dif \omega \, ,
\ee
for a certain function $\omega$ and two arbitrary constants $c_0, c_1$. Therefore, the solution of (\ref{dpdot_dxi}) can be obtained as:
\be \label{A5}
\dot{p}  = c_2 e^{\omega}, \qquad \xi = c_2 e^{\omega} \left[ \zeta_0 + \frac{c_0 e^{-\alpha}}{\beta + c_1}\right]  , 
\ee
where $c_2$ is an arbitrary non-vanishing constant.

\section{\ }
\label{ap-B}

Solving the inverse problem for class C$_{10}$ requires studying the exterior system (\ref{dpdot}-\ref{dxi}-\ref{drho*}) for $\{\dot{p}, \xi, \rho^*\}$ when the integrability conditions hold identically. Let us consider $\{\dot{p}_0 , \xi_0 , \rho^*_0 \}$ a particular solution with $\dot{p}_0 \neq 0$, and let us write $\dot{p}= \phi \dot{p}_0$. Then, equation (\ref{dpdot}) implies that two functions $\bar{\lambda}, \bar{\mu}$ exist such that $\dif \phi= \bar{\lambda} \ell_2  + \bar{\mu} \gamma u$. Moreover, computing $\dif \dot{p} = \dif(\phi \dot{p}_0)$, equation (\ref{dpdot}) becomes:
\be \label{B1}
\dif \dot{p} = \dot{p}_0 \phi \ell_1 + (\lambda  + \phi
\xi_0) \ell_2 + (\mu  + \phi \rho^*_0) \gamma u \, .
\ee
where $\lambda \equiv \bar{\lambda} \dot{p}_0$ and $\mu \equiv \bar{\mu} \dot{p}_0$. Note that $\ell_1 \wedge \ell_2 \wedge u = 2 \gamma a \wedge b \wedge u \neq 0$. Then, identifying the above equation and (\ref{dpdot}), we obtain:
\be \label{B2}
\dot{p} = \phi \dot{p}_0\, , \qquad  \xi=\phi \xi_0 + \lambda\, , \qquad
\rho^* = \phi \rho^*_0 + \mu  \, .
\ee
Moreover, equations  (\ref{dxi}) and (\ref{drho*}) become, respectively,
\begin{equation} \label{B3}
\dif \lambda = \lambda g_1 + \mu g_2 \, , \qquad 
\dif \mu =  \lambda g_3 + \mu g_4 \, ,
\end{equation}
where now the vectors $g_i$ are given by:
\be \label{B4}
\hspace{-15mm} g_1 \equiv \ell_4 - \frac{\xi_0}{\dot{p}_0} \ell_2 \, , \quad g_2 \equiv a - \frac{\xi_0}{\dot{p}_0}\gamma u \, , \quad g_3 \equiv f_2 - \frac{\rho^*_0}{\dot{p}_0} \ell_2\, , \quad g_4 \equiv f_3 - \frac{\rho^*_0}{\dot{p}_0} \gamma u \, .
\ee
The integrability conditions of the system (\ref{B3}) are given by (\ref{dosr}), and they hold identically provided that the exterior system (\ref{dpdot}-\ref{dxi}-\ref{drho*}) for $\{\dot{p}, \xi, \rho^*\}$ fulfills its integrability conditions. Thus, the system (\ref{B3}) meets the conditions of the one considered in \ref{ap-A}. Then, we can obtain $\lambda$ and $\mu$ as:
\be
\lambda  = c_2 e^{\omega}, \qquad \mu = c_2 e^{\omega} \left[ \zeta_0 + \frac{c_0 e^{-\alpha}}{\beta + c_1}\right]  , 
\ee
where the functions $\{\alpha, \beta, \omega\}$ are defined by the conditions (\ref{A3}-\ref{A4}), and where $\zeta_0$ is a particular solution of the Riccati-like equation $\dif \zeta = g_3 + \zeta (g_4 -g_1) - \zeta^2 g_2$. Then, expressions (\ref{B2}) provide the general solution of the exterior system (\ref{dpdot}-\ref{dxi}-\ref{drho*}), where the function $\phi$ meets $\dif \phi= (\lambda/\dot{p}_0) \ell_2  + (\mu/\dot{p}_0) \gamma u$.


\ack 
We are grateful to B Coll for having instilled in us the conceptions that motivate this work and other related research. This work has been supported by the Spanish Ministerio de Ciencia, Innovaci\'on y Universidades, Projects PID2019-109753GB-C21/AEI/10.13039/501100011033,
and the Generalitat Valenciana Project AICO/2020/125. S.M. acknowledges financial support from the Generalitat Valenciana (grant CIACIF/2021/028).


\end{document}